\documentclass[12pt]{article}

\usepackage{tabularx}
\usepackage{array}
\usepackage{booktabs}
\usepackage{multirow}
\usepackage{graphicx}
\usepackage{natbib}

\title{Information content based model for the  topological properties of the gene  regulatory network of {\it Escherichia coli }}

\date{21 November 2009}

\author{Berkin Malko\c c$^1$, Duygu Balcan$^2$ and Ay\c se Erzan$^{3,}$\footnote{Permanent address: Department of Physics, Faculty of Sciences and Letters, Istanbul Technical University, Maslak 34469, Istanbul, Turkey} $^,$\footnote{Corresponding author: erzan@itu.edu.tr}\\\\
$^1${\it Department of Physics, Faculty of Sciences and Letters,} \\{\it Istanbul Technical University, Maslak 34469, Istanbul, Turkey}\\
$^2${\it Center for Complex Networks and Systems Research,} \\{\it School of Informatics and Computing, Indiana University,} \\{\it Bloomington, IN 47408, USA}\\
$^3${\it Department of Physics, Faculty of Sciences and Letters,} \\{\it Akdeniz University, 07058 Antalya, Turkey}}

\begin{document}

\maketitle

{\center{\bf Abstract\\}}
\noindent
{\small
Gene regulatory networks (GRN) are being studied with increasingly precise quantitative tools and can provide a testing ground for ideas regarding the emergence and evolution of complex biological networks.
We analyze the global statistical properties of the transcriptional regulatory network of the prokaryote {\it Escherichia coli}, identifying each operon with a node of the network.
We propose a null model for this network using the content-based  approach 
applied earlier to  the eukaryote {\it Saccharomyces cerevisiae}.~\citep{Balcan2} Random sequences that represent promoter regions and binding sequences are associated with the nodes. The length distributions of these sequences are extracted from the relevant databases.  The network is constructed by testing for the occurrence of binding sequences within the  promoter regions. The ensemble of emergent networks yields an exponentially decaying in-degree distribution and a putative power law dependence for the out-degree distribution with a flat tail, in agreement with the data. The clustering coefficient, degree-degree correlation, rich club coefficient and $k$-core visualization all agree qualitatively with the empirical network to an extent not yet achieved by any other computational model, to our knowledge. The significant statistical differences can point the way to further research into non-adaptive and adaptive processes in the evolution of the {\it E. coli} GRN.}

\section{Introduction}

Complex biological systems such as tran\-scrip\-tional reg\-u\-la\-tory net\-works~\citep{Ecoli_GRN_motifs,Samal},  protein--protein in\-ter\-ac\-tion net\-works~\citep{Mirny} or metabolic networks~\citep{Jeong} all require the satisfaction of  physical and  chemical constraints between  pairs of molecules. There is a growing body of knowledge  regarding specific  protein-DNA or protein-protein interactions and metabolic pathways. Nevertheless, it would be hopeless  at this stage to try to   predict the universal~\citep{Bergman2004} features or the global architecture  of the gene regulatory network or the proteomic network on this  basis.  On the other hand, rudimentary forms of many complex structures we observe in biology can arise spontaneously, given the combinatoric profusion of possible ways in which the simple building blocks, such as nucleotides, amino acids, etc., can be associated with each other. 
~\citep{Dawkinsa,Dawkinsb,Kauffman} In particular, given a sufficiently long sequence (such as the genome), and a set of complex biological molecules (such as the proteins), it is very likely that some of them will have affinities for certain subsequences of the genome and bind them, giving rise to an interaction network.

We propose that important 
insights could be gained into the emergence of biological networks by employing null models making use of the 
combinatoric properties of random sequences~\citep{Kim}.  Comparison with real biological data could enable us to distinguish 
between {\it i) }generic properties of such networks, {\it ii)} features 
that could spontaneously evolve under the kinetics of duplication and 
divergence~\citep{modelgenome6,Sengun,Lynch} and {\it iii)} those 
features that must have clearly evolved in response to specific selection 
pressures (see, e.g.,~\citet{Kashtan} and references therein). Such a null model for the global statistics of a biological network would not aim to model it on a node to node basis, but would provide a much more appropriate starting point for further elaboration than would a ``classical" random network \citep{Erdos1959,Erdos1960}. 

In previous work~\citep{Balcan1,Mungan} we have  demonstrated that any collection of random sequences of varying lengths with a matching rule between them naturally gives rise to a  complex network.  In a recent  paper~\citep{Balcan2} it was shown that, the statistical properties of the transcriptional gene regulatory network (GRN) of the eukaryote {\it Saccharomyces cereviciae} 
(yeast)~\citep{Yeastract} could be predicted by such a null model, with
random sequences representing the binding sites of the transcription factors and the promoter regions. The information content of the binding sequences, extracted from their probability matrices~\citep{Yeastract}, was used to determine their effective length distribution. 
The only free parameter employed in this model was the exponent of the long tailed power law  distribution~\citep{Provata1999} exhibited by the promoter region lengths~\citep{Harbison}. This exponent was chosen to obtain the best fit of the model networks to the empirical GRN. In fact, the model was not very sensitive to the precise value of this exponent.

It should be mentioned that similar ideas of sequence matching have been employed earlier in constructing model genomes and interaction networks~\citep{modelgenome5,modelgenome2,modelgenome1,modelgenome4,modelgenome6,modelgenome3}. However, in these models no attempt was made to take into account the variability in the specificity of the interactions and the results were not very realistic.

Taking the content-based network approach of~\citet{Balcan2} we here propose a model which is able to predict many of the global statistical properties of the GRN of {\it Escherichia coli}, on the basis of the distribution of the degree of specificity of the binding sequences/sites
~\citep{Sengupta1}. If the high degree of agreement between the GRN of yeast and the content-based model is not due to pure chance, we should be able to demonstrate that the model also captures the characteristics of a prokaryotic GRN that is known to be organized somewhat differently. 

\begin{figure}
\includegraphics[width=1.0\textwidth]{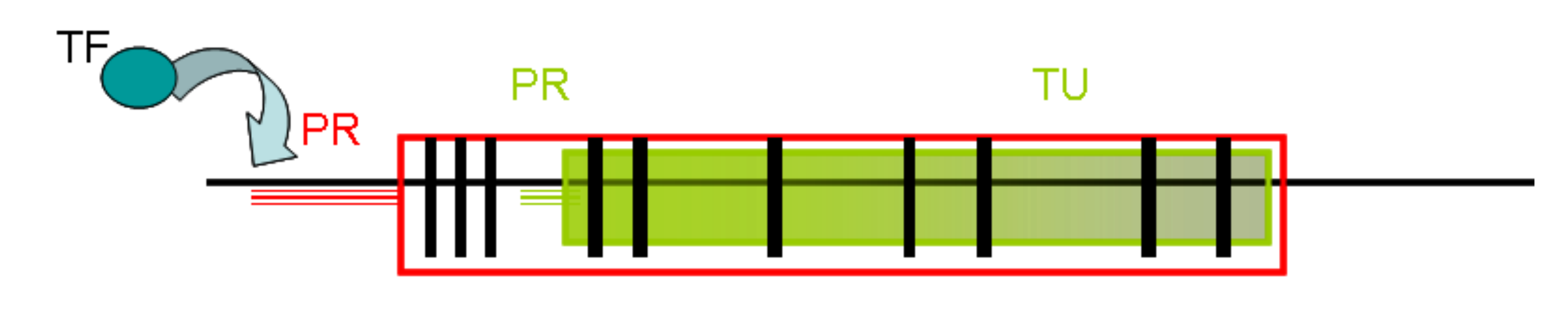}
\caption[]{{\bf Hierarchical organization of the {\it E. coli} genome}.
Shown are an operon 
(big box), which constitutes a transcription unit (TU) in itself, and a smaller box embedded in it, which is 
another TU. Promoter 
regions (PRs), shown here as striped horizontal bars, are attached to both the whole operon and the TU embedded 
in it.  The vertical bars indicate different genes within the TUs. A blob on the left hand side represents a 
transcription factor which may bind a binding site (bs) within one of the PRs, initiating the transcription of the 
TU associated with that PR. The drawing is not to scale.}
\label{Operon-picture}
\end{figure}

In prokaryotes, in contradistinction to eukaryotes, one encounters a more 
complex, hierarchical organization of promoter regions and groups of genes which they 
regulate.~\citep{Cell,Provata1999,Ecoli_TF,Ecoli_GRN_motifs,Okuda,Ecoli_GRN_Salgado,Warren}
Genes which are  to be co-regulated typically  occur in 
tandem. A group of one or more genes which are very 
often (but not necessarily always) transcribed together into one mRNA is called a ``Transcription Unit" (TU). A maximal series of genes that can be transcribed into an mRNA, organized into one or more tandem or overlapping TUs, are termed an ``operon." 
(see Fig.\ref{Operon-picture}) The operon, and the TUs that it may contain, are (generally) preceded by promoter regions (PRs).
As in the case of eukaryotes, special proteins called transcription factors (TFs) have high affinities for certain sequences, called ``binding sequences," or somewhat misleadingly, ``binding sites" (bs) within these 
PRs.~\footnote{The terminology here is far from uniform. Certain authors, e.g., \citet{Cell}, refer to the DNA sequence bound by the RNA polymerase as a ``promoter," and the region in which the binding sequences of the transcription factors regulating a given TU are to be found, is termed a ``gene control region" or a ``regulatory region." On the other hand, \citet{Harbison}, \citet{laessig}, \citet{Provata1999} among many others use the term in the way we have defined it here.} 
The binding of a bs by a TF may facilitate or suppress the transcription of all the genes in the associated TU, thereby regulating their expression. 

In the next section we present our information-content based null-model for the {\it E. coli} GRN. In Section 3, we describe the biological information which determines the model parameters. In Section 4 we outline our simulation results which we obtain by generating an ensemble of realizations of the model GRN, compute the topological properties of these different realizations and compare them with the empirical network which we extract from the RegulonDB v6.0~\citep{RegulonDB}. Section 5 will be devoted to a discussion of our findings.

\section{The model}

To build the model network, we choose a certain number of nodes, each representing an operon.  For each node, we pick 
from an empirically determined distribution (see Fig.\ref{PRnumber}), the 
number of 
TUs to be associated with that operon. This will determine the number of random 
PR sequences which we will assign to the node. The lengths of these PR sequences 
will be chosen from the empirical PR length distribution (see Section 3 
and Fig.~\ref{PRlengths}).

We then randomly choose an  empirically determined percentage (see Table~\ref{parameters}) of operons which will incorporate TF coding genes. 
Each TF-coding node will be assigned one or more binding sequences its TF will recognize. The number of bss per TF obeys the empirical distribution in Table~\ref{parameters}.
The length of each distinct bs will be independently chosen from the bs length distribution (see Fig.~\ref{effective-bs-length-dist}) determined (see next section and Appendix A) from the information content of the known binding sequences.  

Finally, we consider each TF-coding operon (with at least one bs, by definition), and ask if any of these sequences are contained as a subsequence within any of the PRs assigned to any of the operons.  A directed edge is then drawn from the operon coding the TF to each of the operons whose PRs contain the bs associated with this TF. Self-interactions are included in this scheme. Nodes connected by directed edges going both ways are considered to be connected by one bi-directional edge. The resulting directed network is one realization of our model GRN.

\subsection{Information content and specificity of the connections}
Since we require an exact match for each connection, the number of characters in the bs determines the specificity of the interaction. In a real genome,  the consensus bs represents a number of slightly different sequences to which the same TF binds. Therefore we identify the length of a model binding sequence with the effective length computed~\citep{Balcan2} from the probability matrices of the set of similar sequences recognized by a given TF.~\citep{RegulonDB,Stormo_PWM,Siggia,SwissRegulon,Stormo} (see Appendix A).

We define~\citep{Balcan2} the effective information content of a consensus bs as its Shannon information~\citep{Shannon,Avery2003} {\em relative} to a random sequence,
\begin{equation}
I_m= \sum_{i=1}^{l_m} \sum_{j=1}^4  p^{(m)}_{ij} \ln p^{(m)}_{ij} + l_m \ln 4 \;\;.
\label{shannon}
\end{equation}
Here $l_m$ is the length of the $m$th bs and the elements of the probability matrix, $p^{(m)}_{ij}$, are the probabilities for encountering the $j$th 
nucleotide (from among A,T,C,G) at the $i$th site, computed over the different instances of this bs within the different PRs.  The last term comes from subtracting off the information  content of a random sequence of length $l_m$, with equal probabilities for the four letters. Note that any other choice of the background distribution would have just shifted the lengths by a constant amount per position (see Appendix A).

The binding sequences and the PRs will be represented in Boolean characters (see Appendix A). The ``bit," having the values of 0 and 1, is universally accepted as the basic unit of coded information.   We define the effective lengths of the binding sequences represented in Boolean characters to be, 
\begin{equation}
l^{\rm eff}_m = [I_m/\ln 2]\;\;, \label{shannon2}
\label{efflength}
\end{equation} 
where the square brackets indicate integer part. 
The probability matrices are frequently reported in the data bases on the 
assumption that they factorize site-wise, which is most probably not the 
case~\citep{laessig,additivity,barkai,Sengupta,Okuda}, and the effective bitwise lengths are probably slightly overestimated as a result (see Section 3).

\subsection{Connection probabilities}
Within our model, the probability of  connection between two nodes can be estimated analytically. The probability of encountering a Boolean bs sequence of length $l$, associated with one node,  within the PR sequence (of length $k\ge l$) of a second node, is given to a very good approximation 
(see~\citet{Mungan}, Fig. 2) by, 
\begin{equation}
p(l,k) \simeq 1-\left( 1- 2^{-l}\right)^{k-l+1}\;\;. \label{probability}
\end{equation}
For $l\gg 1$, this can be further approximated by $p(l,k) \sim (k-l+1)2^{-l}$. Thus the effective length distribution of the  bs is highly relevant to the topological properties of the resulting interaction matrix, with the connection probabilities depending exponentially on the bs lengths, and only linearly on the length of the PR.

The resulting network can be considered as a superposition of random networks, with connection probabilities which depend on the properties of different classes of nodes, here labelled by the lengths of the bs (if any) and  PR sequences associated with them. For analytical results on how such a superposition can actually give power law or exponential distributions, the reader should consult~\citet{Mungan} and \citet{Balcan_Chaos}. It should be noted that different distributions of sequence lengths give rise to markedly different behavior. The interaction network of a set of random sequences with an exponential length distribution was solved analytically by~\citet{Mungan}, and shown to exhibit an out-degree distribution with two different scaling regimes and a non-monotonic in-degree distribution with a Gaussian tail. 

\section{Extracting the model parameters from biological data}

\subsection{The GRN at the operon level}
As stated above, to analyze and model the GRN at the operon level, we 
identify the nodes of the network with the operons. A TF coded by any one of the genes belonging to any TU within an operon contributes 
one or more out-going edges to the node associated with that operon.  Conversely, any TF which binds 
 a bs contained within a promoter region (PR) associated with a node will contribute 
an in-coming edge to that node, regardless of whether it regulates the operon itself, or a TU within the operon.

\begin{figure}
\includegraphics[width=1.0\textwidth]{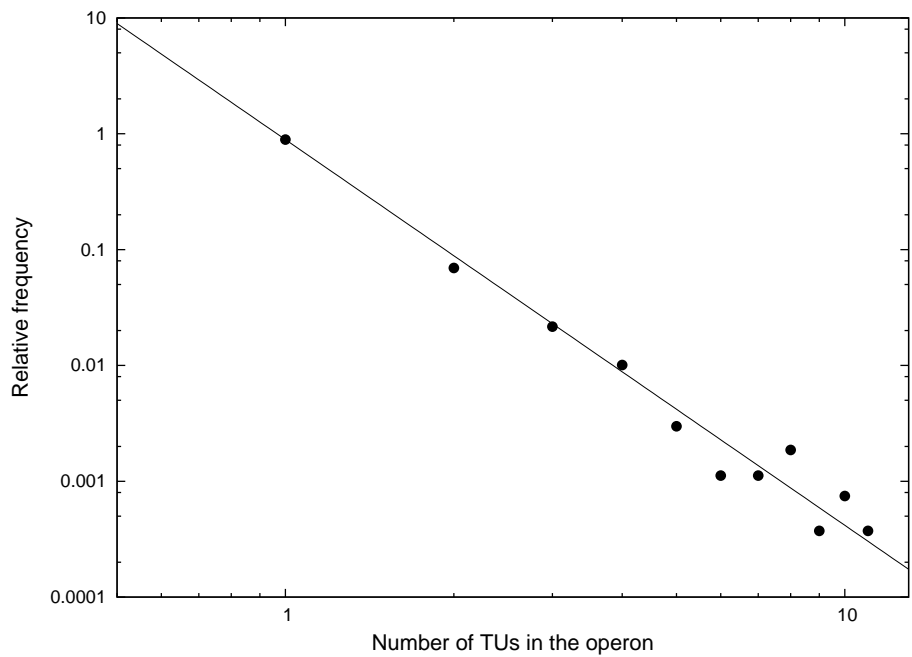}
\caption[]{{\bf Transcription units per operon}. Distribution of the number of transcription units (TUs) per 
operon for the {\it E. coli} genome, 
extracted from the RegulonDB v6.0~\citep{RegulonDB}. The straight line is the power law $x^{-\nu}$, with $\nu=3.333 
\pm 0.045 $.}
\label{PRnumber}
\end{figure}

In the RegulonDB v.6.0
the 
number of TF coding operons are 159 out of a total of 2684, or about $5.9 \%$ of all operons.  The {\it E. coli}  genome is reported to have only one operon which codes two TFs, all the rest code one or zero. There are a couple of TFs which are complexes formed out of two proteins coded by genes in two different operons but we have neglected this fine detail. We picked $5.9 \%$ of the nodes on the average and  assigned one binding sequence to each, indicating that they code  candidate TFs. 
The distribution of the number 
of PRs (effectively, the number of TUs)  associated with each operon is determined from the RegulonDB and is 
shown in Fig.~\ref{PRnumber}.

\subsection{Determining the PR lengths}
Prokaryotes have a very high proportion of ``coding'' to 
intergenic material in their genomes, in comparison to eukaryotes.
In {\it E. coli} coding material constitutes about $89 \%$ of the whole genome, while in {\it S. cereviciae} the ratio is just about inverted. 
The number of genes, on the other hand, are comparable in the two organisms; as a consequence, the average intergenic distance is much smaller on the prokaryotic genome, and distributed exponentially.~\citep{Provata1999}

\begin{figure}
\includegraphics[width=1.0\textwidth]{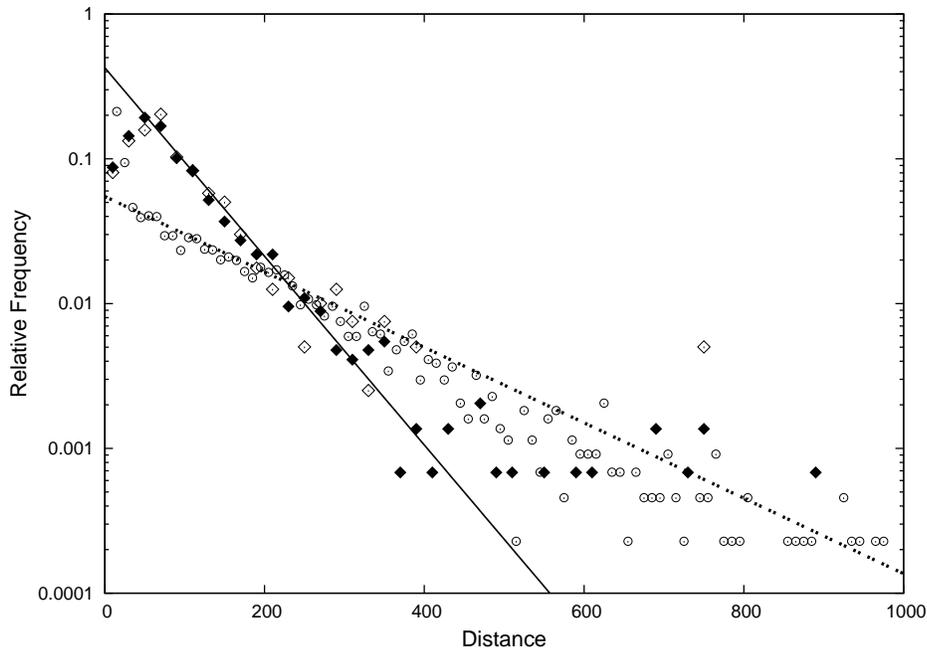}
\caption[]{{\bf Where do the binding sites occur?} The distribution of intergenic distances~\citep{intergenic} in base 
pairs (circles) and the distances from the centers of the binding sequences (bsc) to the start 
 codons of the operons  or transcription units (TUs) which they regulate (from the RegulonDB v6.0.).  The latter distances are shown as filled diamonds for operons,  empty diamonds for TUs within an operon. The  absolute values of the distances, which may be negative (upstream) or positive (downstream) were taken.
 The distributions  were fitted to 
 exponential functions $\sim  \exp (-b\,l)$, omitting the first three points and roughly ten outlying data points (out of about $100$) with distances up to $ l\simeq 2500$. The $b$ values are given in Table~\ref{parameters}.
}
\label{PRlengths}
\end{figure}

There is no clear cut prescription for the determination of the lengths of the PRs. 
%
We decided to focus on the distances (in number of base pairs, with the 
absolute value taken) between the start codons of the TUs and  the binding site centers (bsc) recognized by the TFs regulating them  (see Fig.~\ref{PRlengths}). We believe this quantity is most clearly indicative of the length of the region in which the bs could possibly occur.  The PR length distribution found in this way is not conditional on whether the regulated TU is buried inside an operon, or is located right at the beginning. 
By contrast, the intergenic distances found from the EcoGene database~\citep{intergenic}, also displayed in  Fig.~\ref{PRlengths}, are larger for inter-operonic pairs of genes than they are for intra-operonic pairs~\citep{Okuda}. The continuous line in Fig.~\ref{PRlengths}, fitted to the relative frequency of bsc-to-start-codon distances taken from the RegulonDB, corresponds to an exponential distribution  $p_{\rm PR}(l) \sim  \exp(-b \, l)$ with 
$b=0.0152\pm 0.0007$ (see Table~\ref{parameters}).  In performing the actual simulations, the PR lengths were randomly selected from the empirical distribution of all the bsc-to-start-codon distances shown as diamonds in Fig.~\ref{PRlengths}, uniformly shifted upwards by 9 base pairs to allow the shortest PRs to accommodate binding sequences of typical length.  The mean of this shifted distribution is 91 bp, with a few datum points at distances as large as 2500 bps. The absolute range of the distribution is comparable with that for {\it S.cereviciae}, although there the distribution decays only as a 
power law~\citep{Provata1999,Balcan2}. 

\subsection{Probability matrices of the binding sequences}
The most important problem was in determining the effective lengths (see Eqs.~\ref{shannon},~\ref{shannon2}) of the binding sequences.
We have analyzed the {\it E. coli} data starting from version 5.6 of the RegulonDB
and subsequently updated our data with versions 5.7, 5.8 and 6.0. 
In the successive updates of the data base, the most telling difference was in the small but extremely important (see Eq.~\ref{probability}) upward shift  of the minimum effective bs length appearing  in  RegulonDB v6.0.   
Moreover, inspection of the sets of sequences used for the generation of certain probability matrices in the RegulonDB revealed that, if the sequences were clustered into several distinct sets (rather than being considered as variants of the same bs) they could be better aligned. This would result in several probability matrices with fewer columns but with larger matrix elements, leading to larger relative information content (Eq.~\ref{shannon}) for several {\it distinct} binding sequences.~\citep{Fu} 

The literature search for an alternative source for weight matrices yielded the  SwissRegulon~\citep{SwissRegulon} and PRODORIC (v2.0)~\citep{Prodoric}. 
 The effective bs length distributions obtained via Eqs.(~\ref{shannon},~\ref{shannon2}) from  these three  data bases are displayed in Fig.\ref{effective-bs-length-dist}.  Data was  binned into intervals of size three. 

The weight matrices (see Appendix A) quoted in the Swis\-s\-Reg\-u\-lon database were obtained 
by~\citet{SwissRegulon} by re-clustering and re-aligning the binding site 
data from RegulonDB, using the clustering algorithm "PROCSE" 
of~\citet{Nimwegen}.  In Fig.\ref{effective-bs-length-dist}, the computed length distribution in bits is 
indicated by the open circles. This distribution is clearly much smoother than the others. The lower limit of its range  
agrees with that of the 
RegulonDB v6.0 and it has a mean of 20 bits. We have superposed on this 
set of 
points a truncated Poisson distribution with the same mean, normalized 
over their finite range. It can be seen that most of the points fall right 
on the curve, which interpolates over the gaps in the data.  In our 
simulations we have randomly chosen our bs lengths from this smoothed, 
Poissonian distribution.

We have found from the SwissRegulon~\citep{SwissRegulon} data that the number of binding 
sequences per TF obeys the distribution given in Table~\ref{parameters}.  
Each bs contributes a single length to the empirical length distribution, 
regardless of the number of other binding sites a TF may have. In 
constructing the model genome each bs length is drawn independently from 
the truncated Poisson distribution shown in 
Fig.\ref{effective-bs-length-dist}.

The parameters for the GRN network of {\it E. coli} are shown in  
Table~\ref{parameters}.

\section{Simulation results}

We chose the size of the model and empirical networks to be comparable, even though the absolute size of the network can be normalized away for all of the statistical graph properties discussed below, except for the $k$-core analysis.  
The empirical network has 683 nodes (out of 2684) that have at least one edge connecting them to some other node. For the model networks we started with 2684 nodes as in the empirical network; the number of connected nodes range between 575 and 1372, with a mean of 982 and standard deviation 162.
The model networks have on the order of 2000 edges, whereas the empirical network has about 1300. 

In our simulations we randomly  pick $5.9 \%$  of the nodes to be candidates for TF coding nodes;  however only about half of them actually connect (see Table~\ref{parameters}) and we  end up with about half the number of TFs as the empirical network.~\footnote{We have repeated the  calculations with twice the number of nodes that are TF candidates, ending up with approximately  the empirical number of TFs which actually bind other nodes. The statistical distributions  characterizing the network, normalized as they are by the size of the network in each case,  remain the same. There is  a slight reduction in the scatter due to the larger network. Data  is available upon request.}    

In order to study the statistical properties of our model, we have performed two sets of simulations.  In the first, we have computed the topological properties of 100 realizations of the model. Below, in Figs.5-11,
we display the scatter plots we thus obtained.  The properties computed 
for the empirical network are superposed on these simulation results. We have also performed $k$-core analysis of the empirical and model networks. 
Finally, we have  extracted statistics of the motifs~\citep{Alon2,Alon3} encountered in both the empirical and model networks.  

In the second set of simulations, we have randomized the empirical and model networks while keeping  the in- and out- degrees of each node fixed. Comparisons of the topological properties of randomized versions of the empirical and model networks are available in Appendix C. Not surprisingly, the empirical network moves closer to  the null model  under randomization, while the 100 randomized versions of one randomly picked model network generate another, statistically identical realization of the original ensemble.  As remarked below, the clustering coefficient and motif statistics of the empirical graph are most strongly affected by the randomization, while the degree-degree correlation function is almost left invariant. 

The simulation code for generating the adjacency matrices and computing the graph properties was written in C$^{++}$. A reasonably annotated version is available upon request. The random number generator we used was a C$^{++}$ implementation by Richard 
Wagner\footnote{http://www-personal.umich.edu/$\sim$wagnerr/MersenneTwister.html} 
of the Mersenne Twister~\citep{MersenneTwister}.

\subsection{Degree distributions}
A basic tool of graph analysis is the degree distribution of the nodes 
\citep{BA,Dorogovtsev}. The degree of a node is the total number of nodes to which it 
is connected, by one or more directed or undirected edges (see Appendix B).

In Fig.~\ref{degree-dist} we present the results for the degree 
distribution $p(k)$, on a log-log plot.  The emerging picture is rather 
similar to that of the GRN of yeast~\citep{Balcan2}.  It is very 
gratifying that here too, the empirical data points (red disks) fall right 
on top of the scatter of points from 100 independent realizations of the 
model.  For each realization we have picked the relevant sequence lengths, 
TF numbers and bs numbers, from the appropriate distributions and 
generated the random sequences independently.  In 
Fig.~\ref{degree-dist_mean} we have plotted the averages over the 100 
realizations. The error bars correspond to one standard deviation in each 
case.  In this semi-log plot, one may discern that the initial part of the 
$p(k)$ curve is exponential in both the model and empirical networks; the 
difference between the slopes of the fitted curves is somewhat in excess 
of the error bars.(see Table~\ref{comparison})

By plotting the in- and out-degree distributions separately one can see 
that the small degree part of the distribution in Fig.~\ref{degree-dist} 
comes essentially from the in-degrees, while the larger degrees are 
contributed by the out-degrees.  In Fig.~\ref{In-degree-dist-2panel}, the 
semi-log plot for the in-degree, and in Fig.~\ref{Out-degree-dist-2panel}, 
the 
log-log plot for the out-degree distributions, one has strong qualitative 
agreement between the model and empirical networks, with, however, a 
quantitative difference in both the average in-degree per node and the 
power in the initial, scaling range of the out-degree. (also see 
Table~\ref{comparison}.) In the clustering of the model points on the far 
right side of the out-degree distribution, we see a faint remnant of the 
discrete peaks which would be there for a much larger 
network~\citep{Mungan}. Within the model, these peaks arise from the 
shortest bs sequences which are most frequently to be encountered in the 
longer PRs.  As one goes to smaller degrees, the peaks merge and give rise 
to a continuous distribution.

We  find that the degree distribution of the empirical gene regulatory network of {\it E. coli}, as well as that of yeast~\citep{Balcan2}, are much richer than than so far suspected. 
There were early claims~\citep{Barabasi_Naturegenetics,Bergman2004,Dobrin2004,Vazquez2004}  that gene regulatory networks were scale free, with the degree distribution decaying as a universal power law,  $p(k)\sim k^{-\gamma}$, where  $\gamma \simeq 2$, with, perhaps an exponential cutoff. 
For this operon-level analysis we find no evidence for the  power law with an exponential cutoff claimed for the overall degree distribution, and we are not aware of convincing arguments indicating that a process of preferential attachment~\citep{BAmodel} is in operation. On the other hand, our model reproduces  the exponential decay of the in-degree distribution, also noted by~\citet{Guelzim2002} for yeast. For ease of comparison, it may be useful to supply some numbers.  The putative power law behavior of the out-degree distribution has an exponent of $\gamma \simeq 0.3$, over a very limited range of about one decade.   For the empirical network, a comparable fit within an interval of again about a decade yields 0.4 (see Table~\ref{comparison}).  Note that both these numbers are much smaller than 3, expected for the ``preferential attachment" model.

\subsection{Higher order correlations}
Other quantities of interest, which reflect higher order correlations 
between the nodes than just pair-wise connectivity, are the clustering 
coefficient $C(k)$ as a function of the 
degree~\citep{bollobas,BA,Dorogovtsev}, the correlations between the 
degrees of neighboring nodes~\citep{kk-correlation_colizza} and the 
rich-club coefficient~\citep{rich-club,rich-club_colizza}. These 
quantities are defined in Appendix B. Before calculating these three 
quantities, we have removed the self-interactions from both the empirical 
and model networks.  It should be noted that the empirical network has 93 self-interactions, while the model networks have between 0 or 1 self interaction per realization. Since the bs and PR sequences associated with the nodes have been generated independently, this null model does not incorporate the abundance of self-regulatory interactions in the prokaryotic genome~\citep{Lynch}.

The clustering coefficient $C(k)$ is shown in Fig.~\ref{clusteringcoeff}.  
The data points for the {\it E. coli} network follow the same qualitative 
trend as those for the model network, however they are systematically 
shifted to higher values, more markedly so for small $k$ values.  It is 
clear on the log-log plot that the curves followed by both the empirical and the model data points deviate downwards from a straight line and therefore the decay is faster than a power law, contrary to previous claims to this effect \citet{Vazquez2004}.

The simulation results for the rich-club coefficient $r(k)$ (see Fig.~\ref{richclub}) show a more pronounced non-monotonicity where the empirical network  displays a shoulder. There is a shift to higher values in the high-degree end, indicating a greater incidence of inter-connections between high-degree nodes than expected on the basis of uncorrelated binding sequences and PRs. (Both of these  effects also show up in the motif statistics~\citep{Alon2,Alon3}, and it will be  further discussed below.) \citet{rich-club_colizza} find that the rich club coefficient displays a monotonic increase with the degree for real-world networks such as the internet, air transportation networks, and scientific collaborations, as well as random graphs and the scale free networks yielded by the  preferential attachment model of~\citet{BAmodel}. It is interesting that the only departure from this behavior is a small non-monotonicity, or shoulder, for the protein-protein interaction network, which is a constraint-satisfaction type network, like the gene regulatory network we are considering here. 

The average degree of nodes that are nearest neighbor to degree-$k$ nodes 
(the so called $k-k$ correlation, or $k_{\rm nn}(k)$), is plotted in 
Fig.~\ref{k-k}. Here again, one has close qualitative agreement between 
the GRN of {\it E. coli} and the set of model networks.  However, in this 
case the data points are shifted downwards by almost a factor of four in 
the small degree region, indicating that the average degree of neighbors 
of low-degree nodes is four times smaller than what one would expect on 
the basis of the model. This fact is also reflected in the $k$-core 
analysis of the network; see next subsection. In the {\it E.coli} 
GRN, the low in-degree nodes are generally those with high out-degrees, 
regulating a large number of TUs, which are not themselves regulating. 
Thus their neighbors will have degrees that are below the  average. In the model, for any TF-coding node
the PR lengths and the length of the bs associated 
with the TF are chosen independently. Therefore, there is no correlation 
between the in-degree and out-degree of a node. 

\subsection{Hierarchical structure}
A different way of analyzing the graph 
properties is the $k$-core analysis~\citep{bollobas}. The iterative method for determining the different layers, or shells, is described in Appendix B.
The visualization~\citep{Hamelin} of the different $k$-shells
is a very concise way to display the 
hierarchical organization of the graph. 
In Fig.\ref{k-core} we show the $k$-core analysis of the GRN of {\it 
E. coli} at the operon level and a representative realization of our 
model network. Both have five shells. For this application we chose the overall fraction of potential regulatory 
nodes (i.e., those coding transcription factors) to be such that, the 
number of actually connected regulatory nodes was equal to the empirical 
number, 159.  (See Table 1 and Table \ref{supp-comparison}.)

The similarity between the the $k$-core visualizations for the empirical and model networks is very close, with both showing a 
very marked hierarchical organization. All the nodes of highest degree (hubs) reside in the 
innermost core of the graph and are highly connected amongst each other.  Nodes of different coreness (residing in different shells) are preferentially connected directly to the innermost core, with this tendency being more pronounced in the model network. This structure has also been found in the {\it E. coli} GRN at the gene level.  See \citet{Ecoli_GRN_motifs}.

In Appendix D we also provide plots of the shell populations and the connectivity between nodes belonging to different shells.  It is instructive  to contrast the $k$-core analysis of the rather similar yeast GRN with that of Barabasi-Albert scale free graphs of the same  size. The latter yield much fewer shells (only three compared to nine for the empirical and model networks) and no hierarchical organization, with the nodes in different shells being connected to each other seemingly at random. (Fig.1, Supporting Text  2, \citet{Balcan2}.)

\subsection{Motif statistics}
Finally let us consider the motif statistics, reported in 
Figs.~\ref{motifsa},~\ref{motifsb}. We have used the motif finder program ``FANMOD"  
(freely available online at 
http://www.minet.uni-jena.de/$\sim$wernicke/motifs/)  developed by~\citet{fanmod}. 
We see that in the model network bi-directional 
edges are totally absent, so that a number of motifs present in the {\it 
E. coli} GRN are simply ruled out. However we may note that although the 
absolute values of the Z-scores for the motifs in a randomly selected 
realization of the model network are smaller than the values encountered 
in the real network, they do consistently have the same sign, i.e., they 
depart from the randomized versions in the same direction as the empirical 
network. These results may be compared with the motif statistics at the gene level reported by ~\citet{Ecoli_GRN_motifs}.

\section{Discussion}

In this paper we have presented a null model for a complex biological system.  We have provided a detailed comparison of the model with the actual biological network, the transcriptional gene regulatory network of {\it E. coli}. We believe this study contributes to an understanding of how and to what extent  such structures might emerge from combinatoric considerations alone.  

Our analysis of the most up to date data on the transcriptional gene regulatory network of {\it E. coli}  shows that the somewhat simplified  picture of scale free graphs~\citep{Barabasi_Naturegenetics,Dobrin2004,Vazquez2004,Bergman2004}, with exponents $2 <\gamma < 3$, and modeled by a ``preferential attachment" growth rule~\citet{BAmodel},  is not applicable for the statistical features of the {\it E. coli} GRN at the operon level.  The success of our combinatoric model lies in its detailed reproduction of "non-universal" details and trends in the statistical features of the empirical network. 

In Section 4 we have shown that the  model network  presented has an exponential decay  over the range where this behavior is exhibited by the in-degree, rather than a power law as claimed by \citet{Barabasi_Naturegenetics}.  
For the out-degree, the putative  power law behavior  over a small interval of about a decade is reproduced, with a comparably small power (see numerical values in Table~\ref{comparison} for ease of comparison). The long flat tail of the out-degree, extending beyond the small scaling region, is not simply noise, as can be seen from analytic computations, as in \citet{Mungan} and  \citet{Balcan_Chaos}, albeit for different sequence length distributions. 
We find that the clustering coefficient does not follow a power law as claimed by \citet{Vazquez2004}, while our model is able to reproduce the form of the variation with the degree. 
The rich club coefficient displays the same overall increase, as well as a marked 
non-monotonicity as a function of $k/\langle k \rangle$, at the same  
values where the 
graph for the empirical network displays a shoulder. The $k$-core analysis and the $k$-shell population distribution as a function of the coreness (see Appendix D) are in agreement with the main features of the empirical network.


It should be noted that the data that goes into  our model network is of two kinds.  
%
{\it i)} The number of operons, TFs, and the number of different bs bound by the same TF. These determine the total size of the 
network and set lower and upper bounds on the total number of edges, but cannot in any 
way lead to even a qualitative prediction regarding the degree 
distributions or the other topological properties of the network.
{\it ii)} The distribution of the information content 
of the connections, an attribute superficially having nothing to do with the 
topology. 
 Our 
model provides a theoretical framework within which the second type of 
data is used to predict the specificity of the connections, and 
thereby the statistics of the network topology. It should be noted that the range of 
possibilities given just the first kind of data are nearly infinite in the 
absence of a model, and therefore even a qualitative agreement between the 
empirical and model networks is an important achievement.

\subsection{Quality of the data and the predictive power of the model}
The second point we would like to make is that the quantitative agreement 
between the empirical GRN and our model networks improved steadily with 
the discovery of larger and larger numbers of regulatory interactions in 
the {\it E. coli} genome.  It can be seen from 
Fig.~\ref{effective-bs-length-dist} that there is rather poor agreement 
between different data bases regarding the effective binary length 
distribution of the binding sequences. The crucial increase in the minimum 
effective information content of the binding sequences reported in the 
successive versions of RegulonDB (starting from 
v5.6~\citep{RegulonDB_old}), and the improved distribution we derived from 
the SwissRegulon data base~\citep{SwissRegulon}, resulted in a radical improvement in the 
agreement between the model networks and the {\it E. coli} GRN.  Thus, we 
may say that given the correct bs and PR length distributions, the model 
is able not just to mimic but to virtually {\em predict} qualitative features of 
the {\it E. coli} GRN as reported in the RegulonDB v6.0.

\subsection{Effects that have been neglected}
A number of possible reasons can be cited for the small but persistent 
difference between the distribution of empirical in-degrees and those 
estimated from the model network.

A high degree of overlap is found between consensus sequences in the 
relatively short promoter regions of {\it E. coli},
leading us to 
conjecture that even if more than one interaction is allowed in principle, 
only one of them will be realized at any given time. 
``Transcriptional interference"~\citep{TI1}, where interference between RNA 
polymerase binding two close-by sites inhibits transcription of one or both of the TUs, 
has recently been studied and modeled.~\citep{TI3,TI2} Such effects can 
have further consequences for the reduction of the actual regulatory 
interactions from those that are possible purely on the basis of 
combinatoric arguments.

Several workers~\citep{mantegna,Provata2004} have claimed that 
correlations within intergenic regions lead to reduced information content 
(and effective bitwise length) of PRs, by several percent. This 
would reduce the real connectivity of the actual networks to below what we 
conjecture on the basis of random PR sequences.

The binding sequences we obtained from the RegulonDB were
slightly anti-correlated on average. We define the average distance per nucleotide between pairs of binding sequences of the same length $l$ as
\begin{equation}
h=1-(1/4) \sum_l  \left[ \vert S_l\vert l \right]^{-1}
\sum_{\mu, \nu \in S_l} \sum_{i=1}^l \sum_{j=1}^4 p^{(\mu)}_{ij} p^{(\nu)}_{ij}
\end{equation}
where $S_l$ is the set of binding sequences of length $l$, and $|S_l|$ is the size of this set; $\mu$ and $\nu$ indicate different binding sequences within such a set, and $p^{(\mu)}_{ij}$ is the probability matrix for this binding sequence. Had the binding sequences been totally random, with the probabilities for the bases A,T,C,G  given by 1/3, 1/3, 1/6,1/6, we would have gotten 0.728 for $h$, whereas from the RegulonDB(v5.7) we found 0.737, i.e., the binding sequences are farther from each other on the average than random sequences, by $1\%$. Thus, the probability for encountering overlapping binding sequences within a random PR is actually lower than had the former been random, but this is a very small effect which is below noise level in the present discussion. (Note that, in the absence of joint probabilities for the occurrence of different nucleotides at given sites of distinct binding sequences, the mutual information between them, constructed from just the probability matrix, is identically zero.)

~\citet{barkai} report that  in those cases where more than one TF is binding a PR region, a lower  specificity is tolerated, i.e., the binding sequences in this region are ``fuzzier." 
This means that the effective lengths of the binding sequences sought out by the same TF may actually vary between different regions of the genome, an effect we have not taken into account in this model. By keeping the effective length of the consensus sequences fixed, independently of the length of the PR in which they are to be sought, we slightly disadvantage the binding probabilities at the shorter PRs compared to what seems to be observed.  This effect, however, is of the same order (and opposite sign) as that which would be induced by the anti-correlation between the binding sequences, an effect which we ignore.  We believe these two small corrections effectively cancel each other out and that we are not in error in neglecting both of them.

\subsection{Evolution of correlations}
Above all, it is necessary to understand that certain features of the empirical network could never be reproduced by such a naive null-model as the one we propose.  We have already mentioned the absence of self-interactions in the model networks, where the protein product of the gene binds its PR and regulates its own transcription.  Besides these, there are highly conserved, very special regulatory sub-graphs which, 
say, include regulatory nodes with extremely large number of connections, even though the binding sequence which they recognize is highly specific, requiring  the satisfaction of a very large number of constraints.
The correlations between connections, embodied especially in the clustering coefficient and the motif statistics~\citep{Alon2,Alon3} of the network, are other features which are not included in our model, where the assignments of all the sequences associated with the nodes are made independently. It is quite possible, that in the course of the evolution of the GRN, certain nodes with a high out-degree, regulating a relatively large number of TUs, were selected from among those having small in-degrees, introducing a negative correlation between these quantities and leading to the observed discrepancies.  

We believe that instead of comparing the empirical motif statistics with those of purely random networks, it is more meaningful to compare them with the present null model. The most striking feature 
of the motif statistics of the {\it E. coli} GRN is the high 
incidence of bi-directional interactions, giving rise to motifs with the 
highest Z-scores that can be seen in Figs.~\ref{motifsa},~\ref{motifsb}. Such bi-directional interactions 
are in fact present in our model but to a much smaller extent than in the 
empirical model. Note, in Table~\ref{comparison}, that the average total 
degree is very slightly less than the sum of the in- and out- degrees for 
the model networks, while it is markedly different from this sum for the 
empirical network.  The simplest, and most likely~~\citep{laessig,Barkai2} mechanism to give rise to such interactions
is the duplication  of TF-coding genes and their promoter regions, a 
feature which is not present in this model, but which can easily be  built in and has already been considered by \citet{Sengun}.
  Another 
feature which could very easily be incorporated into the model is 
homologies between the TFs leading to similarities between binding 
sequences. The high incidence of the feed-forward loop (motif number 
36, a high Z-score motif) and the large rich-club coefficient would be 
accounted for if there were a high overlap between the binding sequences 
of TFs which regulate each other in a cascade. Further work  on the evolution  of content-based model genetic networks by non-adaptive processes 
\citep{Lynch} is in progress.

\subsection{Comparison with the yeast GRN}
It is instructive to compare our findings for {\it E. coli} with those 
 for {\it S. cereviciae} (yeast)~\citep{Balcan2}. The two genomes 
differ most markedly in the distribution of the lengths of the promoter 
regions, with, however a rather similar distribution of effective lengths 
for the binding sites.  Comparing the GRN of {\it E. coli} with that of 
{\it S. cereviciae} one finds great qualitative similarities between the 
two, with an essentially exponential distribution of the in-degree, a 
rather scattered out-degree distribution  suggesting a power-law 
distribution, and clustering coefficients, degree-degree correlations and 
rich-club coefficients that qualitatively look very similar. 

Comparing the {\it E. coli} and yeast~\citep{Balcan2} networks with respect to their 
$k$-core decomposition is also interesting. A sharp difference between the {\it E. coli} and yeast GRNs shows up in 
the shell population distribution as a function of the coreness:  In the 
case of yeast, the shell population decreses linearly with coreness, 
whereas for {\it E. coli} the decrease is exponential.  The model networks 
mimic these respective behaviors perfectly in both cases.
For {\it E. coli}, the very 
tightly hierarchical connectivity of the model network, with edges going 
almost strictly up and down the coreness hierarchy, is disrupted in the 
empirical network of {\it E. coli} to a greater extent than is the case 
for yeast.  Although the exponential growth trend in the connections to the 
high coreness nodes is common to both these organisms, a greater incidence 
of in-shell (transverse) connections are visible in Fig.\ref{k-core} than in 
the corresponding Fig. 2 of \citet{Balcan2}.  This behavior is  
graphically illustrated in Fig. \ref{edge_distribution}, where the empirical 
graph deviates from the exponential growth of the connectivity to higher 
coreness nodes, and shows an excess of connections to nodes of low 
coreness.  This agrees with a well known feature of the 
prokaryotic genome where there is an abundance of small regulatory loops and 
self-regulatory interactions~\citep{Lynch}.

The quantitative agreement between the {\it E. coli} genome and our 
model ne\-tworks is over\-all less than the cor\-re\-spond\-ing agree\-ment found for yeast~\citep{Balcan2}. This could be ascribed to the absence of any fitting parameters in the present study, while in the case of yeast, the (unknown) exponent of the length distributions of the promoter regions was optimized to get the best fits. However, the qualitative dependence of various network features on this number was very weak.  We conjecture that selective pressures on the very compact prokaryotic genome might have caused greater departures from purely  combinatoric features in the {\it E. coli}, than is the case for the yeast genome.

\subsection{Generic features coming from a large number of independent constraints}
In this paper we have constructed a model of the prokaryotic GRN. We should recall that we do not intend to model the GRN on a node to node basis, but only with respect to its global statistical properties.  We have checked whether the number of transcription units (TUs) which a TF regulates is correlated with the information content of its binding sequence, and found no correlation at all, for any of the data bases used. (No such one-to-one correspondence was found in the case of {\it S. cereviciae} either.~\citep{Balcan2}) Thus the high degree to which our model predictions are borne out points to a phenomenon of a more fundamental nature.~\citep{Hwa}  It seems to imply that the distribution of the specificity of the connections seems to arise independently of the actual lengths of the binding sequences recognized by the TFs, but nevertheless has essentially the same truncated Poisson distribution as the latter. 

The clue to this  convergence lies in considering an arbitrary number $m$ of independent conditions for, say, a genomic interaction to be established.  The probability ${\cal P}_m$ for the satisfaction of all $m$ of these conditions will be a product of the individual probabilities, viz.,
\begin{equation}
{\cal P}_m = \prod_{i=1}^m p_i \;\;.
\end{equation}

\begin{figure}
\includegraphics[width=1.0\textwidth]{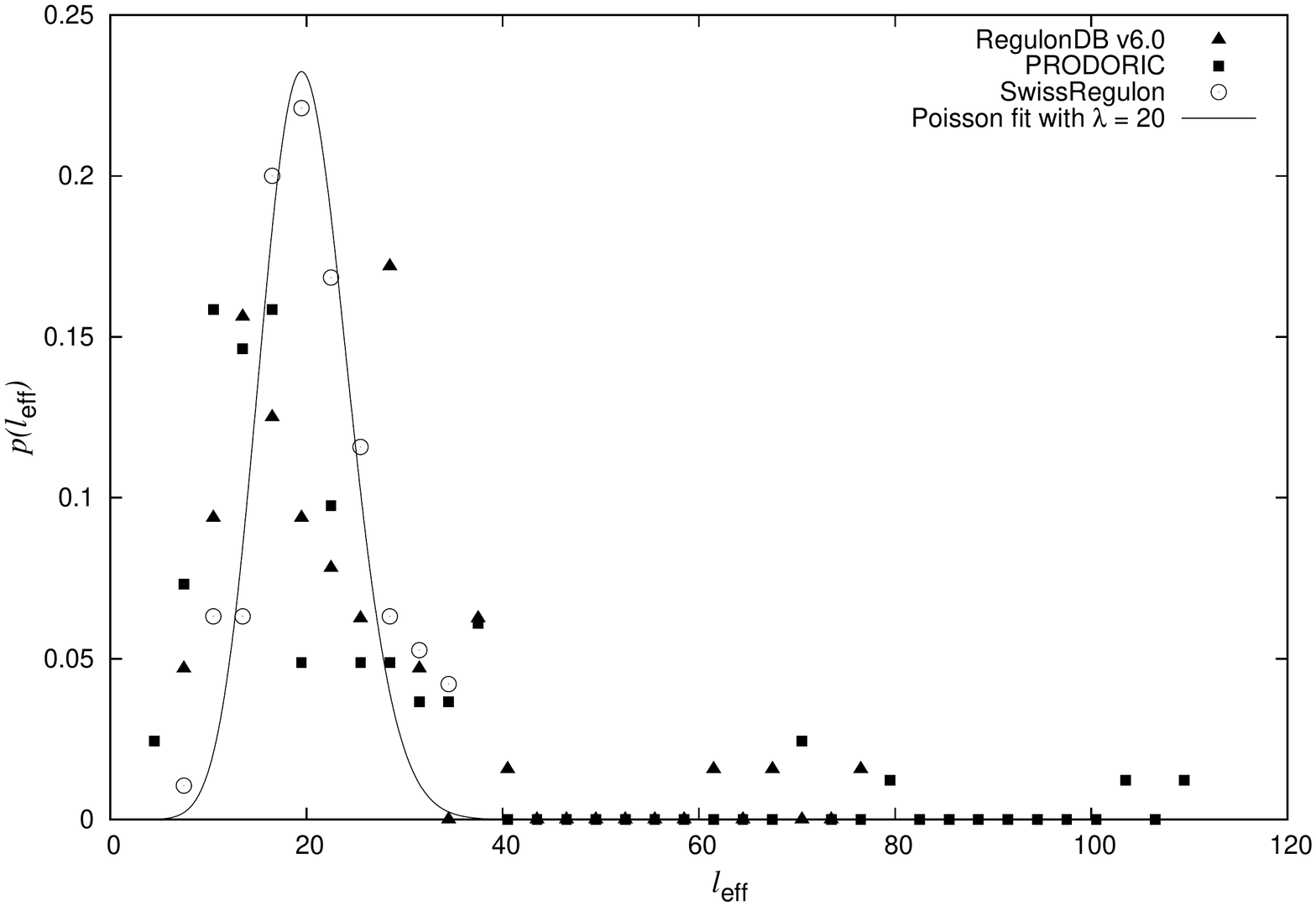}
\caption[]{{\bf Information content of the binding sequences.} Effective binary length distribution for the 
binding sequences of {\it E. coli} extracted from three 
different databases: RegulonDB v6.0~\citep{RegulonDB} (triangles), PRODORIC~\citep{Prodoric} (squares), 
SwissRegulon~\citep{SwissRegulon} (circles).  The solid line is the truncated Poisson distribution with mean 
$= 20$, and normalized over the finite range of the data points represented by circles, as obtained from the 
SwissRegulon data.  The data points fall right on the curve.}
\label{effective-bs-length-dist}
\end{figure}

Each of the probabilities $p_i$ can be expressed as $p_i = 2^{-\alpha_i}$, 
where $\alpha_i = - \ln_2(p_i)$. Thus, ${\cal P}_m = 2^{-\lambda}$ where 
$\lambda= \sum_i^m \alpha_i$.  Even if the $\alpha_i$ are not identically 
distributed, as long as the mean and variance exists for each $\alpha_i$, 
and, for example, Lyapunov's condition (see any standard text on 
probability theory, e.g.,~\citep{Lyapunov}) is fulfilled, we can avail 
ourselves of the central limit theorem, to claim that for $m$ sufficiently 
large, $\lambda$ is Gaussian distributed around $\sum_i^m \langle 
\alpha_i\rangle$ with variance $\sum_i^m \sigma^2_i$, summed over the individual variances. At the level of precision of the fit in 
Fig.~\ref{effective-bs-length-dist}, this distribution 
would be indistinguishable from a Poissonian, especially for  
$\langle \lambda \rangle$ as large as 20, as found here. This argument is in fact very 
general and need not apply only to the establishment of genomic 
interactions. It has to do with the well-known fact that the distribution 
of probabilities for the satisfaction of a large number of independent 
conditions is log-normally distributed.

{\bf Acknowledgements}

It is a pleasure to thank Volkan Sevim 
for a critical reading of our manuscript.  We thank Meltem Sevgi for 
having obtained an early version of the PR-length distribution in the 
initial stages of this work. Berkin Malko\c c acknowledges support from 
the Scientific and Technological Research Institute of Turkey 
(T\"UB\.{I}TAK) National Scholarship Program for PhD Students. Ay\c se 
Erzan would like to acknowledge partial support from the Turkish Academy 
of Sciences.
\vskip 1cm

\pagebreak
\clearpage

\begin{table}
\caption{{\bf Network pa\-ram\-e\-ters and genome data ex\-tracted from the 
Reg\-u\-lonDB~\citep{RegulonDB} and Ecogene~\citep{intergenic}}. The percentage of transcription factors (TFs) 
recognizing $n$ distinct binding sequences has been calculated on the 
basis of the 
60 TFs for which this information is available. The parameter $b$ refers 
to fits to distributions of the form $\exp(-b\,l)$ for the intergenic and 
for binding site center to start codon distances $l$.(see 
Fig.\ref{PRlengths})}
\centering
\begin{tabular}{lccccccc}
\toprule
\multicolumn {5}{ l }{Number of nodes (operons)} & 2684 && \\
\multicolumn{5}{l}{Number of known TFs}  &   159 &&\\
\multicolumn{5}{l}{Percentage of TF coding operons}  & 5.9 & &  \\
\midrule
Candidate PR lengths &&&&\multicolumn{3}{c}{ $b$} & \\
Intergenic &  & 	&	& \multicolumn{3}{c}{$0.00648 \pm 
9\times 10^{-5}$}& \\ 
  bsc-to-start-codon (operon)& &	&	&\multicolumn{3}{c}{$ 0.0156\pm 
0.0003$}&\\ 
  bsc-to-start-codon (TU in op.)& & & & \multicolumn{3}{c}{$ 0.014\pm 
0.001$}& \\ 
  bsc-to-start-codon (op.s and TUs)&& &	&\multicolumn{3}{c} 
{$0.0152\pm 0.0007$}&\\
\midrule
Number $n$ of bs per TF & 1 & 2  & 3 & 4 &5 
 &7 &9 \\
Percentage of TFs with $n$ bs & $76.6 $  & $11.7 $  & $1.7 $ 
 & $3.3 $   & $3.3 $ & $1.7$ & $1.7 $ \\
\bottomrule
\end{tabular}
\label{parameters}
\end{table}

\begin{table}
\caption{{\bf Comparison of the average degrees and degree distributions of the empirical and model networks}. Only those nodes (operons) with non-zero degree have been included in the network statistics. The parameters $\xi$, $\xi_{\rm in}$ refer to the exponential fits to the degree and the in-degree  distributions (see Fig.~\ref{degree-dist_mean} and Fig.~\ref{In-degree-dist-2panel}). The out-degree distribution has a putative power law behavior $k_{\rm out}^{-\gamma}$  for relatively small degrees (see Fig.~\ref{Out-degree-dist-2panel}). These numbers are only provided for ease of quantitative comparison of the model and empirical networks. The power law fits are valid only within a range of about a decade and do not represent any claims that the respective networks are scale free.  See text, Section 4.1.}
\centering
 \begin{tabular}{  l  c  c  }
\toprule
 & {\it E. coli} & {\bf Model}   \\  \cmidrule{2-3} 
Average degree $\langle k \rangle $ & 3.796 &  2.906 \\
$\xi$  & $ 0.46\pm 0.01 $ & $0.35 \pm 0.03$\\
 Average in-degree $\langle k_{\rm in}\rangle$ & 1.977 & 1.454 \\
$\xi_{\rm in}$ & $1.81 \pm 0.09$ & $0.94 \pm 0.03$\\
 Average out-degree $\langle k_{\rm out}\rangle$ & 1.977 & 1.454 \\
$\gamma$ & $0.40\pm 0.01$ & $0.32\pm 0.01$ \\
\bottomrule
\end{tabular}
\label{comparison}
\end{table}

\clearpage

\begin{figure}
\includegraphics[width=1.0\textwidth]{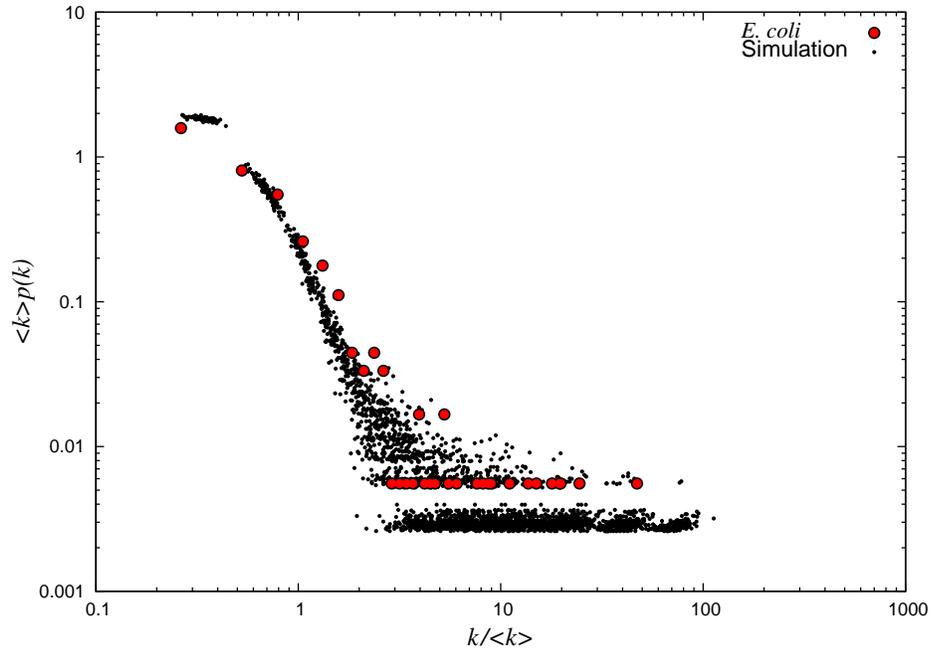}
\caption[]{{\bf Degree distribution.} Plot of the degree distribution of the transcriptional gene regulatory 
network (GRN) of {\it 
E. coli} extracted from the RegulonDB v6.0
(red disks), compared with the scatter plot (black points) 
of the degree distribution of 100 independent realizations of the model network, in which we have included only those nodes with degree greater than zero. To account for the fluctuations in the network size,  the horizontal axis has been 
scaled with the average degree per node $\langle k \rangle$, an extensive quantity. The probability $p(k)$ (the vertical axis) has 
been multiplied by the average degree in anticipation of an exponential fit to the distribution, shown in Fig.~\ref{degree-dist_mean}.
}
\label{degree-dist}
\end{figure}

\begin{figure}
\includegraphics[width=1.0\textwidth]{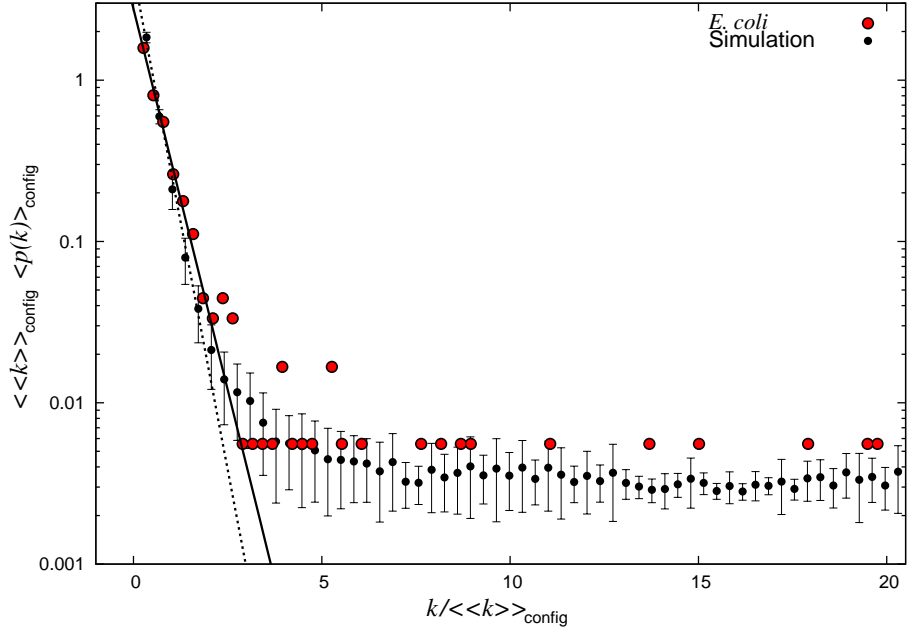}
\caption[]{{\bf Exponential fit to the degree distribution.} Semi-logarithmic plot of the degree distribution 
of the 
transcriptional gene regulatory network (GRN) of {\it E. coli} from the RegulonDB v6.0,
compared 
with the degree distribution of the model network, averaged over 100 
realizations (black discs). The 
configurational average is taken over the set of independent realizations 
of the model network. Error bars stand for one standard deviation. 
Numerical results for the fit to $ \sim \exp(-k/\xi)$ of the initial range 
of the empirical and model distributions are given in Table\ref{comparison}. The vertical axis has been multiplied by the degree averaged over the nodes (and for the model networks, also the realizations) in order to scale away the fit parameters $\xi$.
}
\label{degree-dist_mean}
\end{figure}

\begin{figure}
\includegraphics[width=1.0\textwidth]{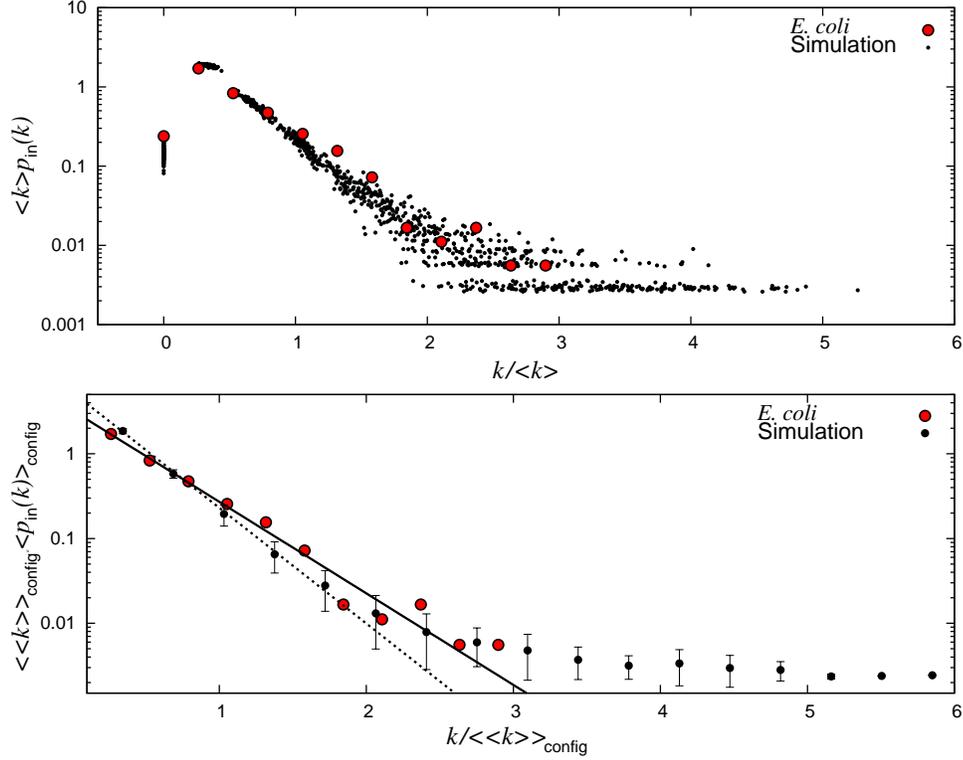}
\vspace{0.5cm}
\caption[]{ {\bf The in-degree distribution.} Semi-logarithmic plot of the in-degree 
distribution of the transcriptional gene regulatory network 
(GRN) of {\it E. coli}, extracted from the RegulonDB v6.0, 
compared with the in-degree 
distribution of the model network.  The lower panel has been averaged over 100 realizations of the model. Error bars stand for one standard deviation. 
Numerical results for the characteristic in-degree, $\xi_{\rm in}$, found from fitting   
$\langle p(k_{\rm in})\rangle_{\rm config} \sim \exp(-k/\xi_{\rm in})$) to the initial ranges of 
the distributions, are given in Table~\ref{comparison}.  The datum point with zero in-degree and the flat tail 
is truncated in the lower panel.}
\label{In-degree-dist-2panel}
\end{figure}

\begin{figure}
\includegraphics[width=1.0\textwidth]{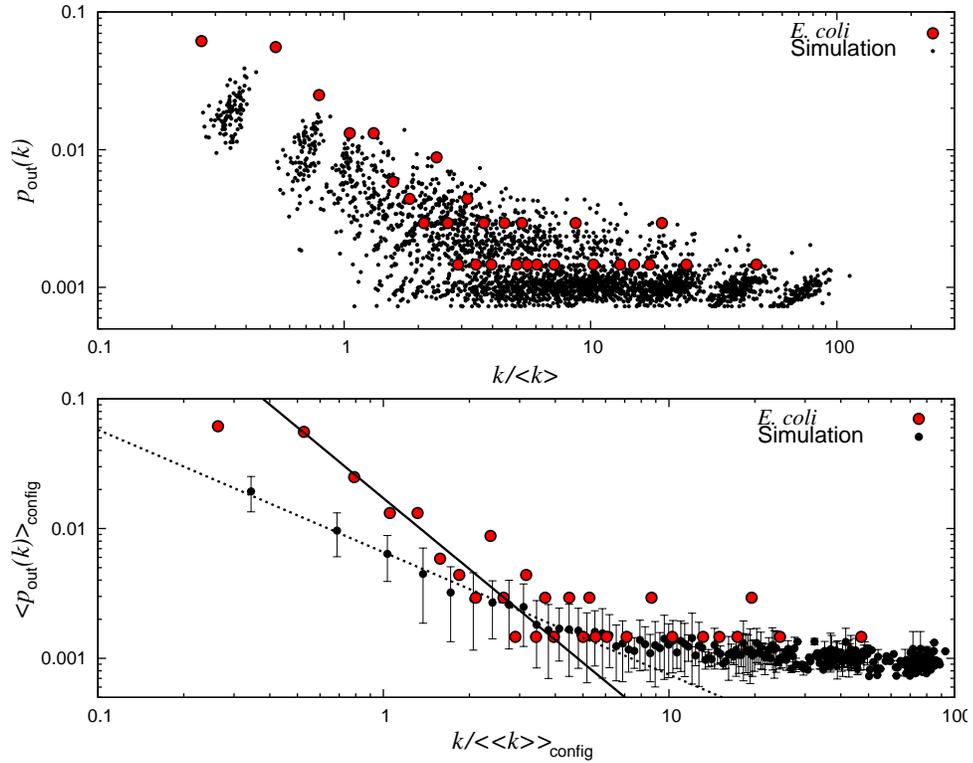}
\vspace{0.5cm}
\caption[]{ {\bf The out-degree distribution.}  Both 
the scatter plot and the averages over 100 realizations of the simulation 
(black discs) are shown, compared with empirical distribution (red discs). 
Error bars in the lower panel stand for one standard deviation. Values of 
the exponents for the fits to $\langle p(k)\rangle_{\rm config} \sim 
k^{-\gamma}$ are given in Table~\ref{comparison}. }
\label{Out-degree-dist-2panel}
\end{figure}

\begin{figure}
\includegraphics[width=1.0\textwidth]{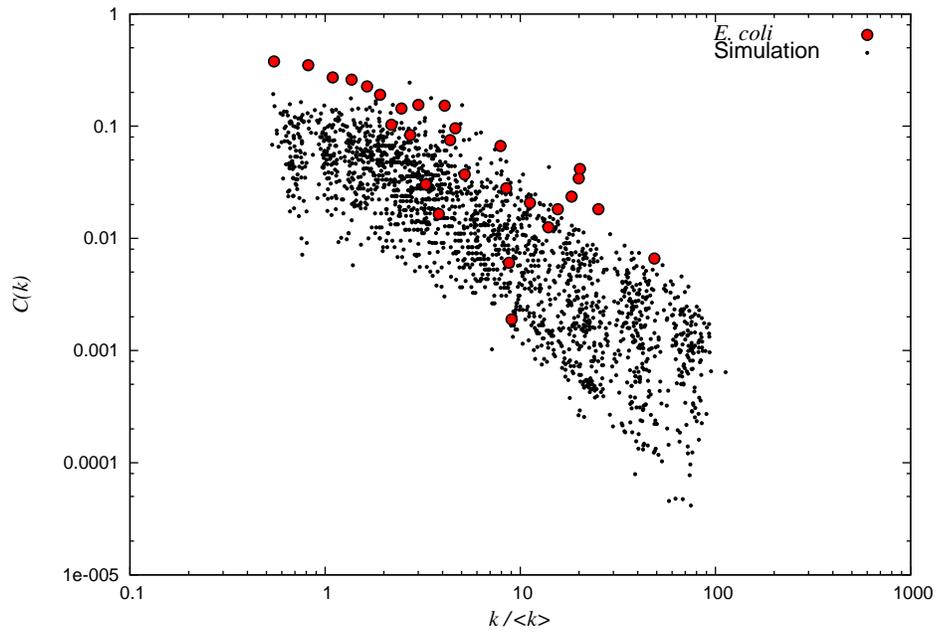}
\caption[]{{\bf Clustering coefficient $C(k)$}.  The {\it E. coli} data from the RegulonDB v6.0
 is shown as red discs. The scatter of black points 
corresponds 
to 100 realizations of the model network.  All self-interactions have been removed from the network before calculating the clustering coefficient.
}
\label{clusteringcoeff}
\end{figure}

\begin{figure}
\includegraphics[width=1.0\textwidth]{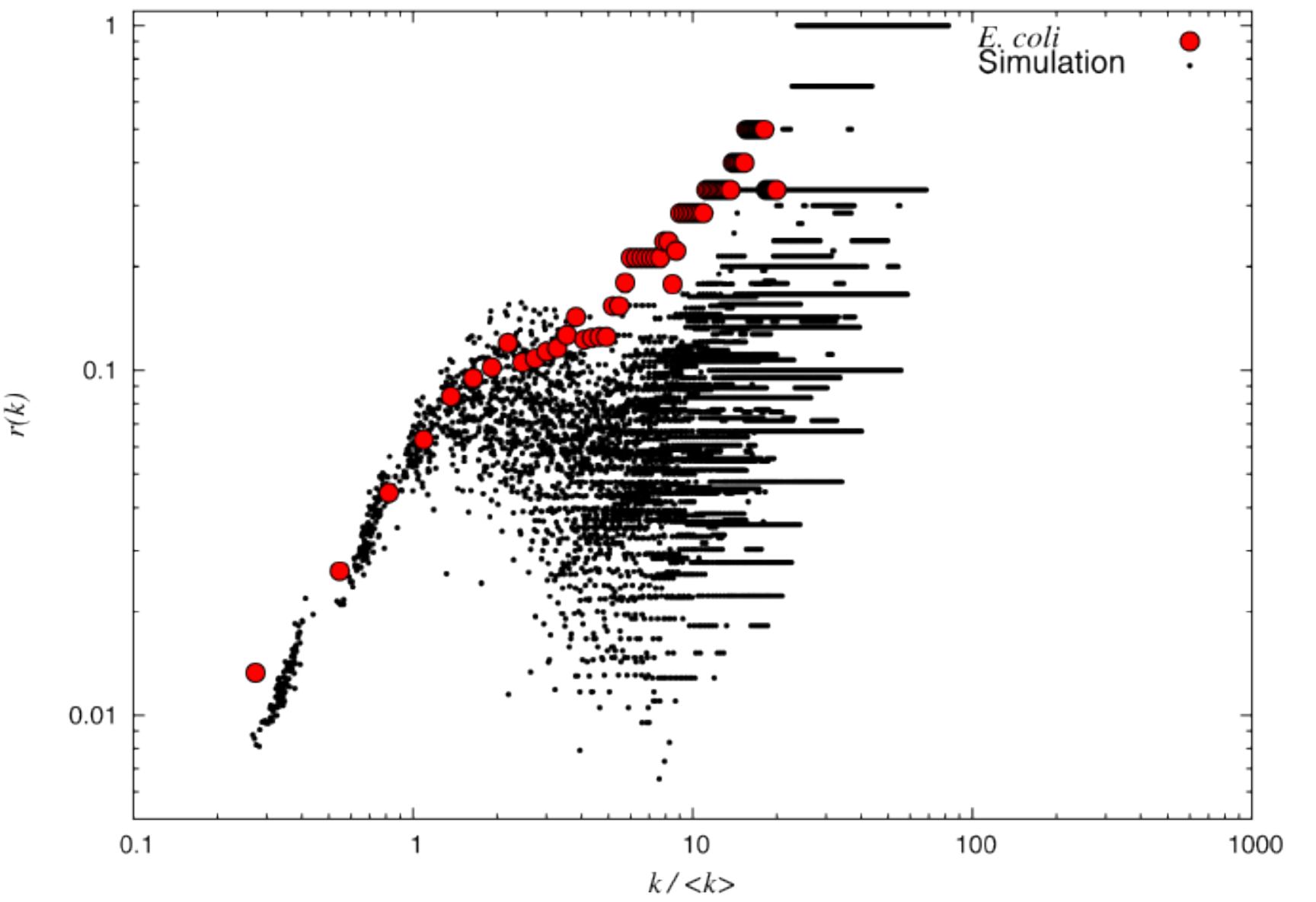}
\caption[]{ {\bf The rich-club coefficient $r(k)$.} The {\it E. coli} data from the RegulonDB v6.0
are shown as red discs. The scatter of black points corresponds 
to 100 realizations 
of the model network. All self-interactions have been removed from the network before calculating $r(k)$.
}
\label{richclub}
\end{figure}

\begin{figure}
\includegraphics[width=1.0\textwidth]{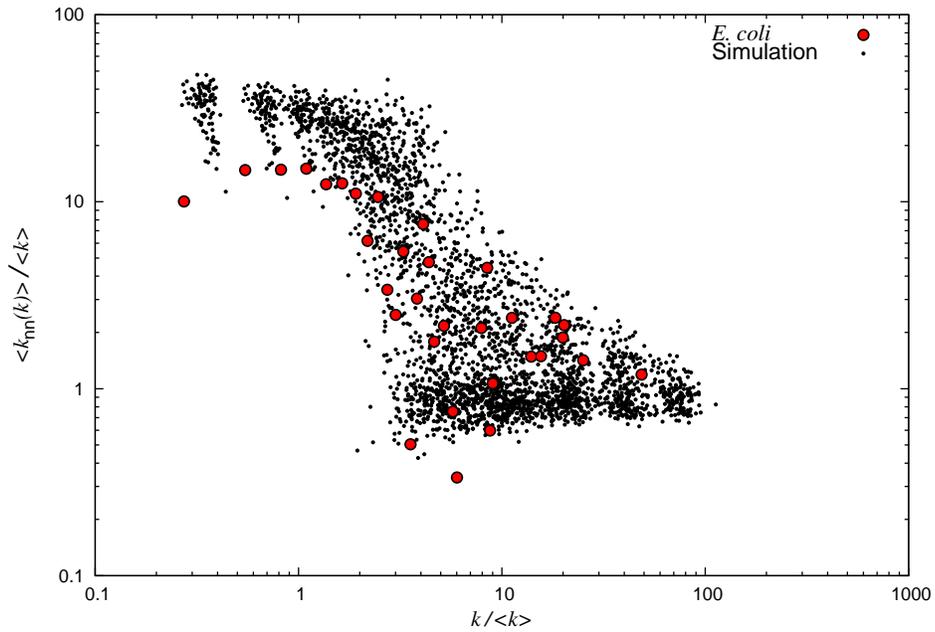}
\caption[]{ {\bf The degree-degree correlation function.} The  expected degree of nodes neighboring a 
degree-$k$-node is denoted by $k_{\rm nn}(k)$. All self-interactions have been removed from the network before calculatingthe correlation function. Since this correlation function is an extensive quantity, both the vertical and horizontal axis have been normalized 
by the average degree. The {\it E. coli} data (red disks) is from the RegulonDB v6.0.
}
\label{k-k}
\end{figure}

\begin{figure}
\includegraphics[width=1.0\textwidth]{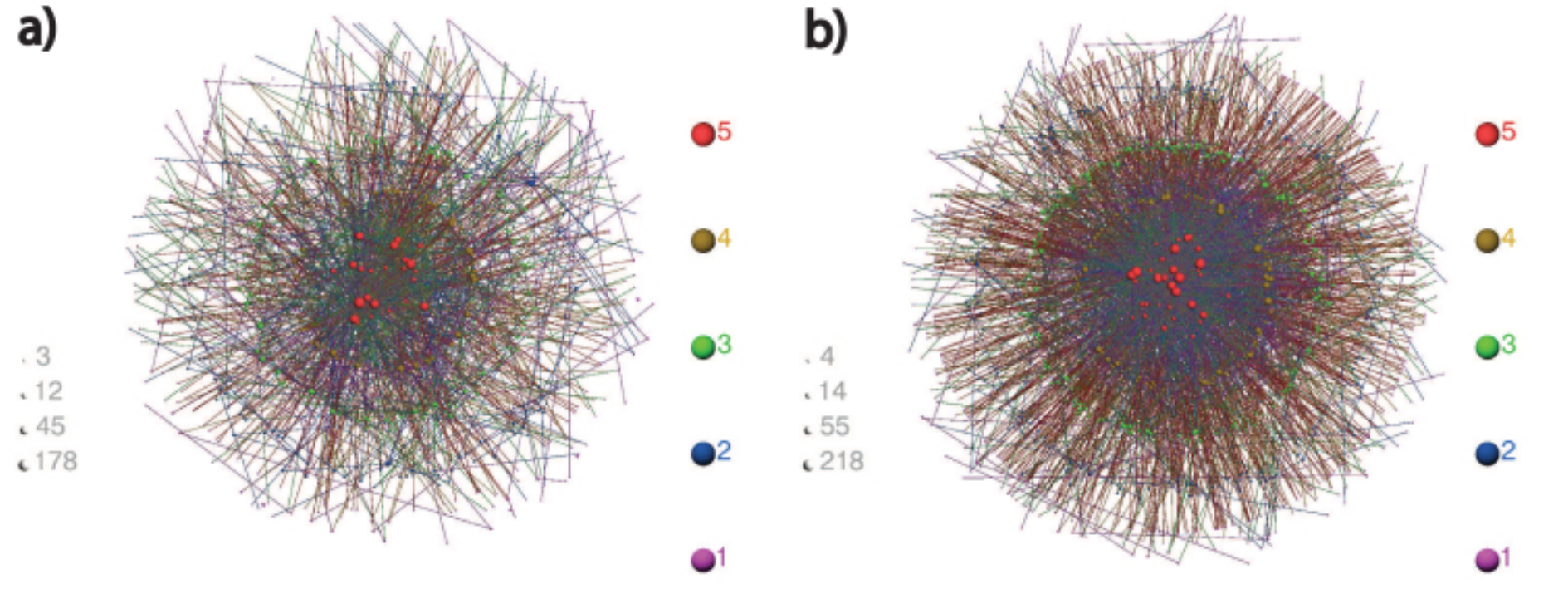}
\caption[]{{{\bf The $k$-core analysis}. (a) The empirical {\it E. coli} gene 
regulatory network from the RegulonDB,
 and  (b) a model network (a typical realization, number 48) from 
our ensemble of model networks.  The $k$-cores have been visualized using the 
visualization tool developed by \citet{Hamelin},   LaNet-vi, which is 
available online at 
(http:$//$xavier.informatics.indiana.edu$/$lanet-vi$/$).
The color code indicated on the right corresponds to the shell number 
(coreness), while the size of each ball is proportional to the degree of 
the corresponding node. A sample of values are given on the left, the last 
one being the largest degree on the network. The thickness of the shells 
corresponds to the spread in the coreness of the nodes to which members of 
a given shell are connected.}
} 
\label{k-core}
\end{figure}

\begin{figure}
\includegraphics[width=1.0\textwidth]{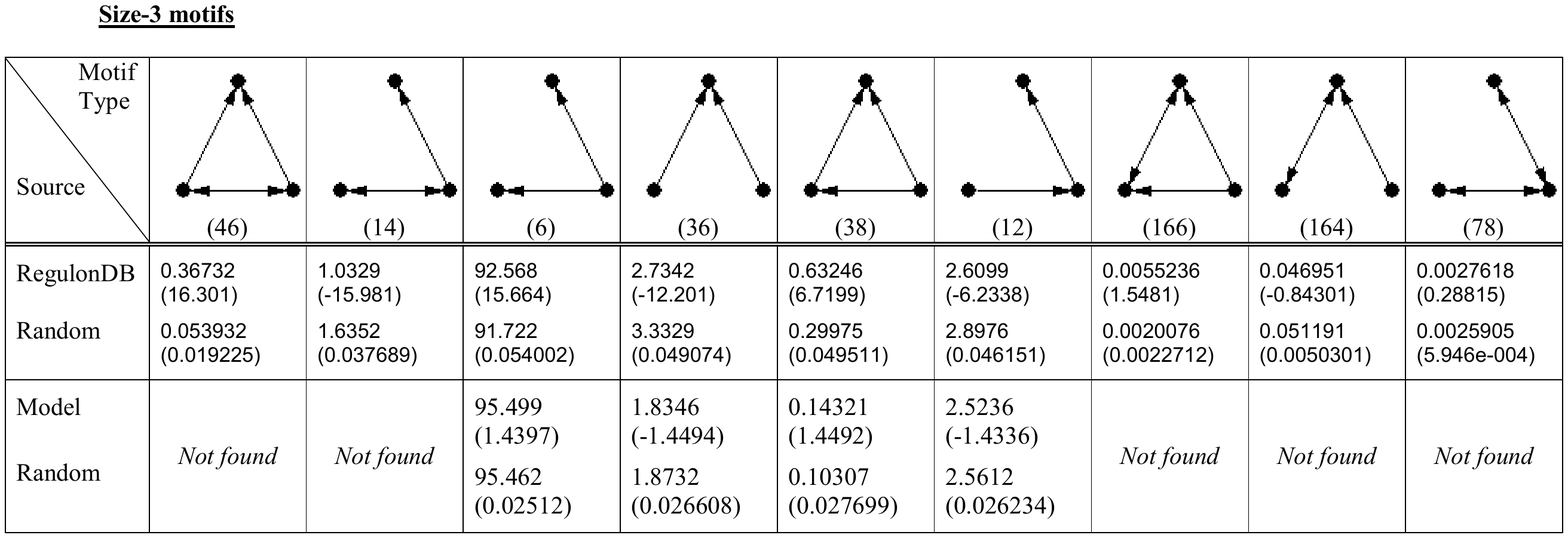}
\caption[]{{\bf Network motifs.} Percentages of size-3 motifs found in the {\it E. coli} network 
generated using data from RegulonDB v6.0
and in one realization of the model. Corresponding values obtained from 1000 randomized networks are also given. Numbers 
in the parentheses for the randomized networks stand for the standard deviations whereas the ones for the original 
network are the Z-scores for that motif. Z-score is defined as the difference between the value for the original 
network and the mean value over 1000 randomizations divided by the standard deviation. Motifs are ordered in 
decreasing value of the Z-score for the RegulonDB network from left to right.   The numbers below the graphs identify the motifs.
}
\label{motifsa}
\end{figure}

\begin{figure}
%
\includegraphics[width=1.0\textwidth]{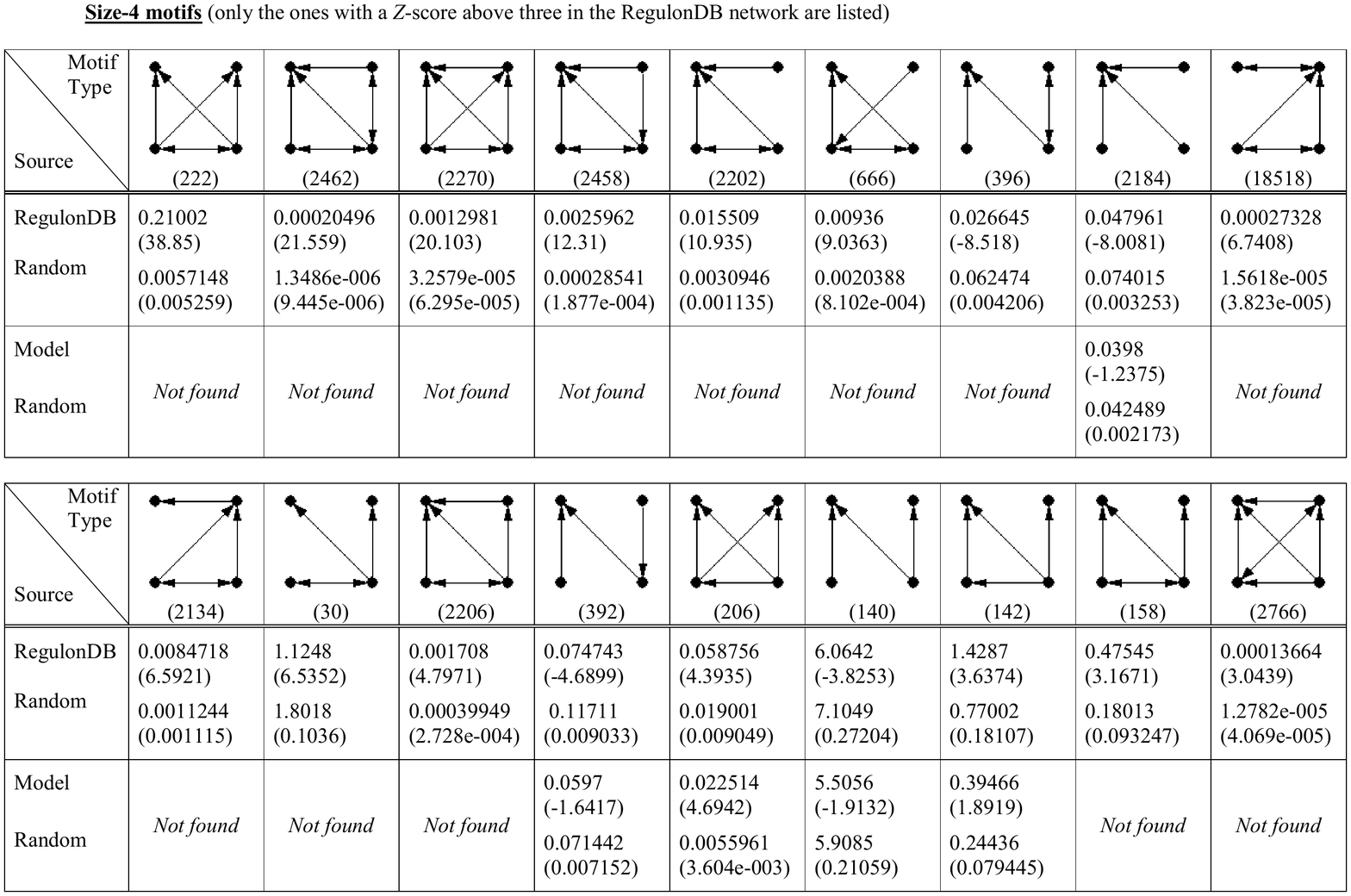}
\caption[]{ {{\bf The statistics for motifs of size 4.}} }
\label{motifsb}
\end{figure}

\clearpage
{\bf Appendix}


\appendix

 \renewcommand{\theequation}{A-\arabic{equation}}
  \setcounter{equation}{0}  

\section{Information content of sequences}

The methodology here closely follows that presented in \citet{Balcan2}. We define the information content of an ensemble of sequences of $N$ letters drawn from an alphabet of $r$ letters 
as~\citep{Shannon,Avery2003},
\begin{equation}
I\equiv \sum_{i=1}^N \sum_{j=1}^r p_{ij}\ln p_{ij} \;\;,
\end{equation}
where $p_{ij}$ is the probability of encountering the $j$th letter in the 
$i$th position.  Note that this is a negative quantity, and therefore is 
sometimes defined with an overall (-) sign out front, in analogy with the 
thermodynamic entropy. It makes sense to subtract from this expression
the information content of a sequence of the same length but with 
letters drawn at random, namely $I^{(0)}\equiv N \sum_{j=1}^r p_{j}^{(0)}\ln p_{j}^{(0)}$, where 
$p_j^{(0)}$ are the ``background"  probabilities of 
the different letters. 
This gives an information content that is 
relative to the random case. We have taken the background 
probabilities to be uniformly equal to $1/r$ in Eq.~\ref{shannon}.

We note that any symbol within an alphabet of $r$ letters can be uniquely 
assigned a Boolean code of length $n$, where $n$ is the first integer 
greater than or equal to $\ln r/ \ln 2$. Any sequence coded in an alphabet 
of $r$ letters can therefore be recoded in 0s and 1s.  Thus the nucleic 
acids of the genomic code, of which there are four, can be uniquely 
represented by 00, 01, 10 and 11, i.e., by four sequences of two bits.  The definition we have chosen for the effective lengths of the binary sequences, Eq.~\ref{shannon2},  is nothing but the number of bits necessary to code a binary sequence with information content equal to that represented by the probability matrix for the consensus sequence. 

The terminology regarding the probability matrices is not  at all uniform.  Some authors prefer to quote frequencies rather than  probabilities, as  in the ``alignment matrices" defined by~\citet{Siggia}. ``Weight matrices"~\citep{additivity}, sometimes called Position Specific Weight 
Matrices~\citep{Stormo_PWM,SwissRegulon} may be used  instead of probability matrices, and they  are defined as  $w^{(m)}_{ij} = \ln p^{(m)}_{ij} -\ln p^{(0)}_j$, where the $p^{(0)}_j$ are the background probabilities for the nucleotides $j$ over the whole genome; $m$ indexes a particular consensus bs. Within this convention the (relative)  information content of a sequence of length $l_m$ is defined as,
\begin{equation}
I_m = \sum_{i=1}^{l_m} \sum_{j=1}^4 p_{ij}^{(m)} w_{ij}^{(m)}\;\;.
\end{equation}
Note that this differs from our definition, in that the subtracted quantity is not the information content of the random series but $\sum_{i=1}^{l_m} \sum_{j=1}^4 p_{ij}^{(m)} \ln p^{(0)}_j$.

\section{Topological characterization of complex networks}

The in-degree of a node is defined as the number of directed edges 
incident upon that node. The out-degree is, conversely, the number of 
directed edges leading out of a node. The (total) degree is the number of 
distinct neighbors of a node, with the neighborhood being established 
either with in- or out-edges, or both.

The clustering coefficient~\citep{bollobas,BA,Dorogovtsev} as a function 
of the degree $k$ is defined as,
\begin{equation}
C(k) = \vert G_k\vert^{-1} \sum_{i\in G_k} \; \sum_{\mu<\nu; \mu,\nu \in \Omega_i}  2e_{i, \mu \nu}/ [k(k-1)] 
\end{equation}
where $G_k$ is the set of nodes of degree $k$, $i$ ranges over the elements of this set, $\vert G_k\vert$ is the size (the number of elements) of this set, $\Omega_i$ is the set of neighbors of the $i$th node, $\mu$ and $\nu$ range over this neighborhood and $e_{i, \mu \nu}$ is either zero or one depending upon whether the $\mu$th and $\nu$th members of the neighborhood of $i$ are disconnected or connected.
	 	
The degree-degree correlation function~\citep{kk-correlation_colizza} is defined as
\begin{equation}
k_{nn}(k) = \sum_{k^\prime} k^\prime p(k\vert k^\prime)
\end{equation}
where $p(k\vert k^\prime)$ is the conditional probability that a node with degree $k$ has a neighbor of degree $k^\prime$.
	 	
The rich-club coefficient~\citep{rich-club,rich-club_colizza}  is the total number   of edges connecting nodes with degree greater than $k$, normalized by the maximum possible number of such connections,
\begin{equation}
r(k)  = 2 e_{>k} / [N_{>k}(N_{>k}-1)]
\end{equation}
where  $N_{>k}$ is the total number of nodes with degree greater than $k$ and $e_{>k}$ is the total number of edges between such nodes.

The $k$-core analysis~\citep{bollobas} of the network into  different layers, or ``shells" is performed via the following iterative 
procedure:  All nodes that are at least of degree 1 will be called the the 
1-core of the graph. The graph may consist of more than one connected 
component.  To start the iteration, all nodes which are connected with one 
edge only are eliminated by severing that edge. The process is repeated 
until no nodes remain which are singly connected to the graph.  What 
remains of the graph is the 2-core, and all nodes outside it are termed 
the 1st shell (although some of them might have had degree greater than 
unity). At the second stage, one searches for nodes that are doubly 
connected to the rest of the graph and removes them together with their 
edges, and the process is repeated until none such are left.  This yields 
the 3-core, and those nodes which have been removed at this stage make up 
the 2-shell. In each $k$-core, the nodes are of degree $\ge k$, and the 
$k$-shell consists of those nodes that belong to the $k$-core but not to 
the $k+1$st core. The ``coreness" of a node is defined as the $k$ 
value of the shell to which it belongs.~\citep{Hamelin} One proceeds as outlined above 
until all nodes are exhausted.  This means that once $k_{\rm max}$ has 
been reached, iteratively removing all nodes with degree $k_{\rm max}$ 
leaves an empty set of nodes.

\section{Randomized versions of the empirical and model networks}

\begin{figure}
\includegraphics[width=1.0\textwidth]{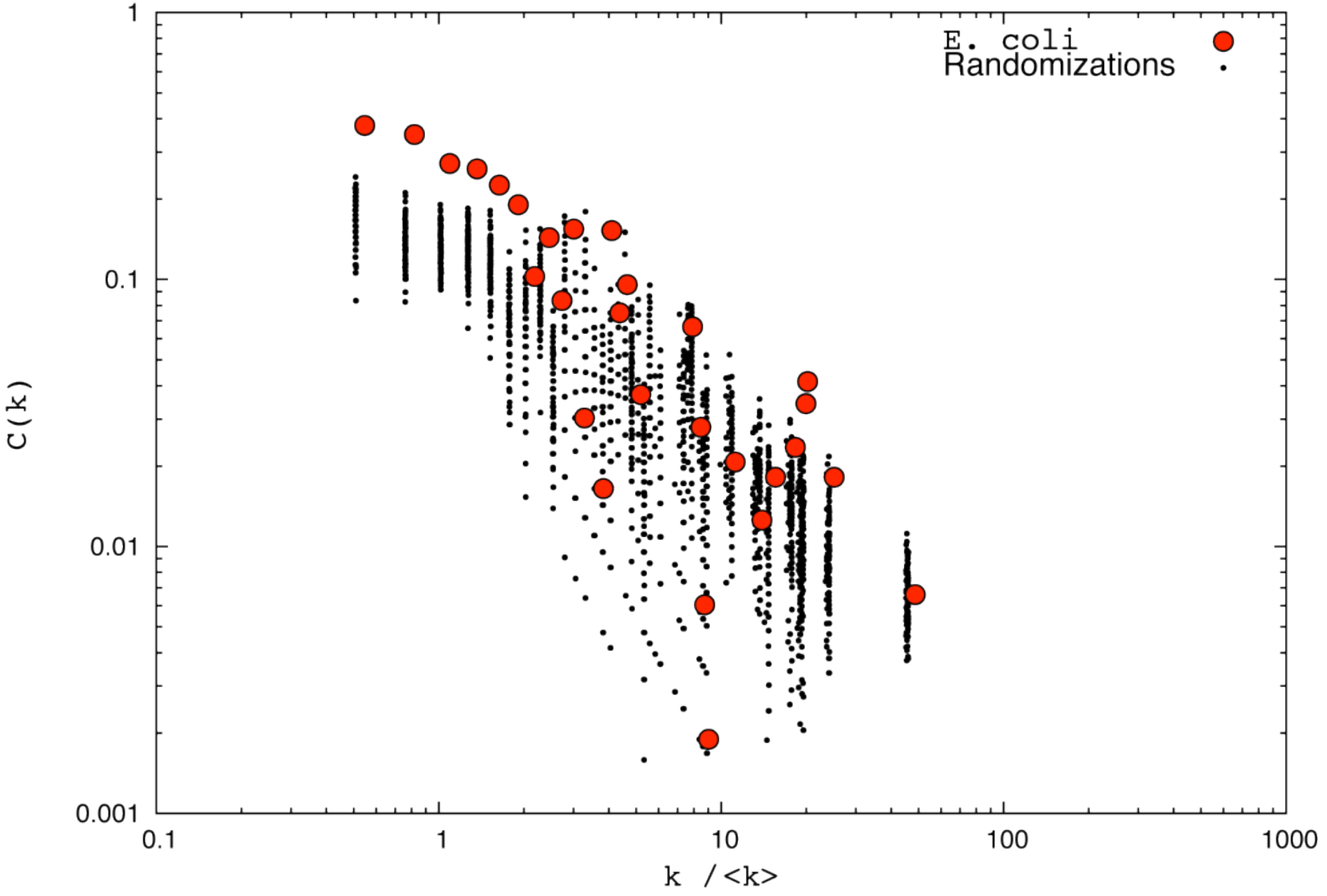}
\caption[]{{\bf Effect of randomization on the clustering coefficient of the {\it E. coli} 
GRN.} The original data points  (red discs) from RegulonDB v6.0~\citep{RegulonDB} are superposed on the 
data points obtained from 100 independent networks, generated by randomly rewiring the edges of the empirical network, keeping the in- and out-degree of each node fixed 
separately.  Rewiring means exchanging either the outgoing or the incoming ends of randomly picked pairs of edges. This operation has been repeated ten times the total number of edges, to get each independent rewired graph.}
\label{CC_random}
\end{figure}

\begin{figure}
\includegraphics[width=1.0\textwidth]{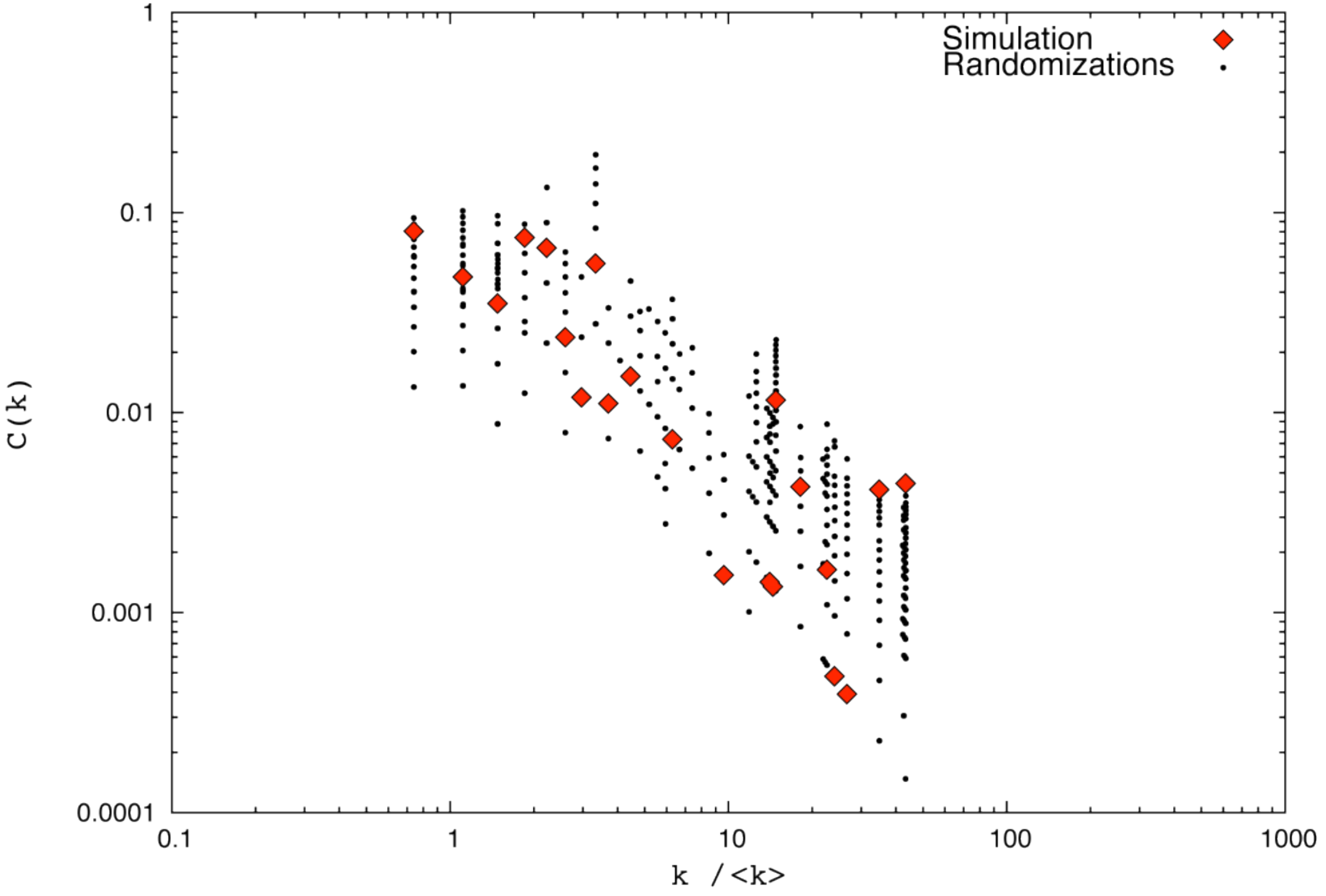}
\caption[]{{\bf Effect of randomization on the model genome clustering coefficient.} The same procedure is 
followed as in 
Fig.~\ref{CC_random}, with the original network (red diamonds) chosen randomly from one 
of the model realizations.}
\label{CC_sim_random}
\end{figure}

\begin{figure}
\includegraphics[width=1.0\textwidth]{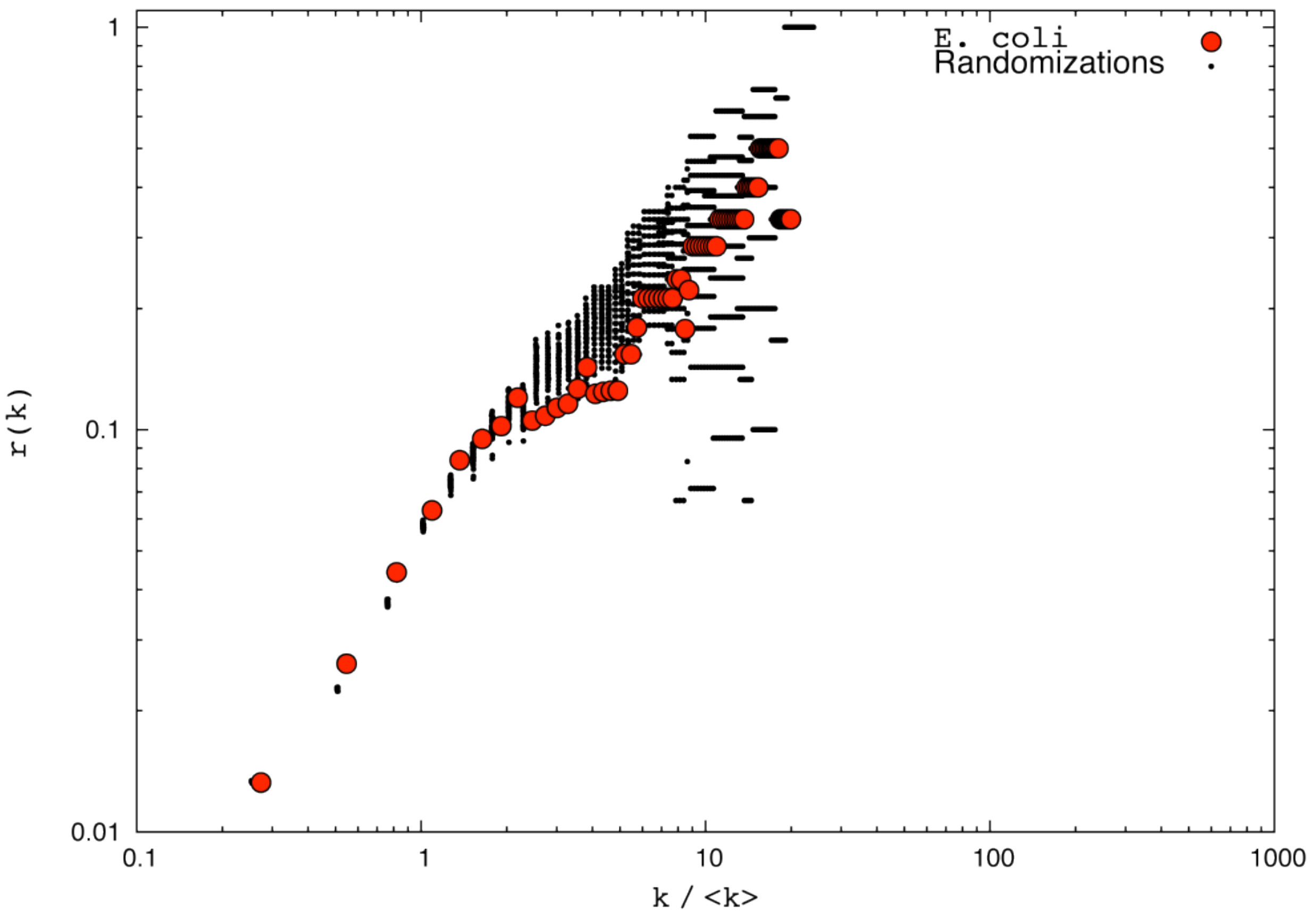}
\caption[]{{\bf Effect of randomization on  the rich-club coefficient, empirical data.} The data for the 
{\it E. coli} 
GRN from RegulonDB v6.0~\citep{RegulonDB} (red discs) is superposed on the data points obtained 
from 100 independent rewirings, keeping the in- and out-degree of each 
node fixed, separately.}
\label{RCC_random}
\end{figure}

\begin{figure}
\includegraphics[width=1.0\textwidth]{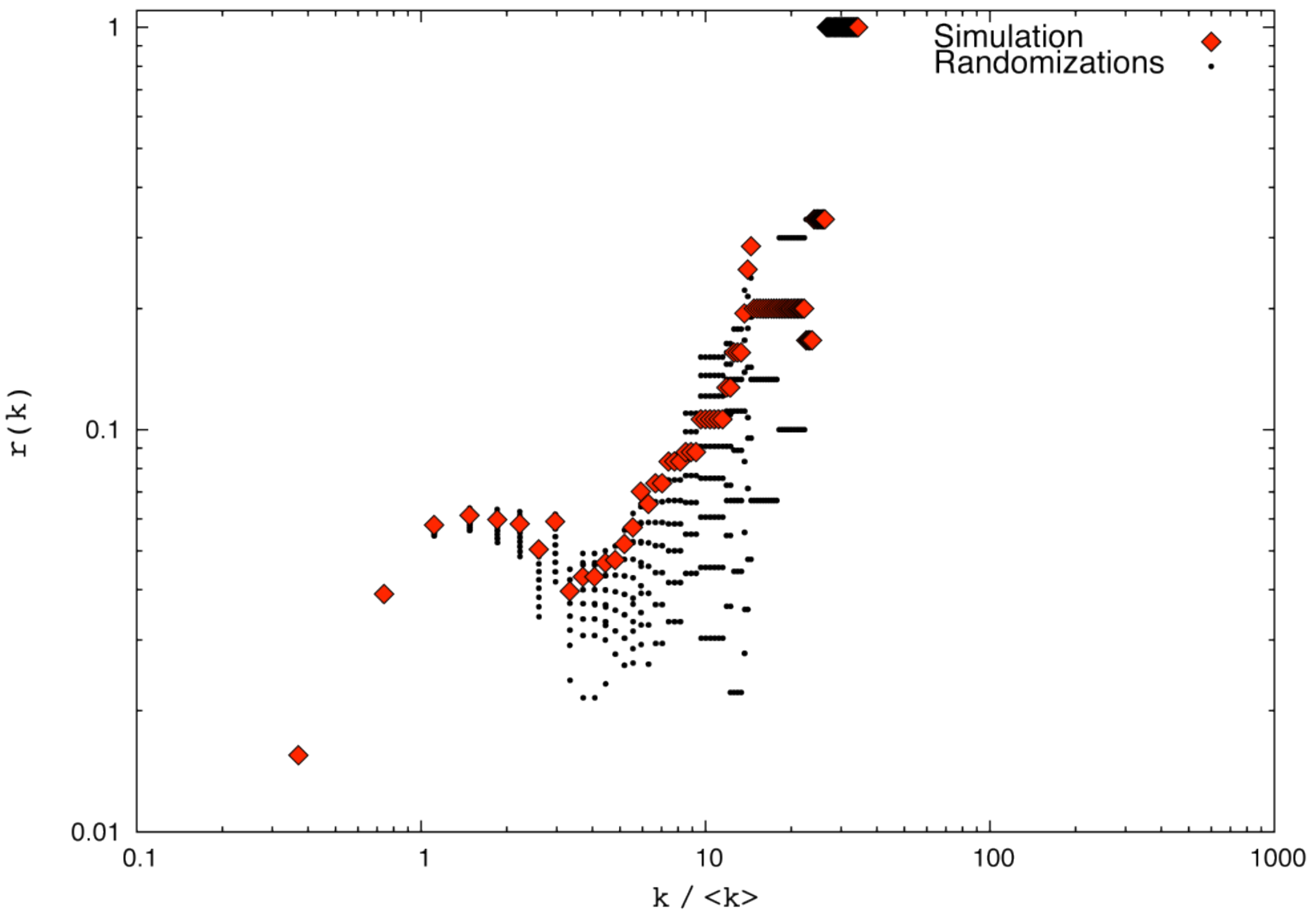}
\caption[]{{\bf Effect of randomization on the model network: rich-club coefficient}. Same as in 
Fig.~\ref{RCC_random}, except that instead of the 
empirical network, a randomly chosen 
model network (red diamonds) has been randomized as described above.}
\label{RCC_sim_random}
\end{figure}

\begin{figure}
\includegraphics[width=1.0\textwidth]{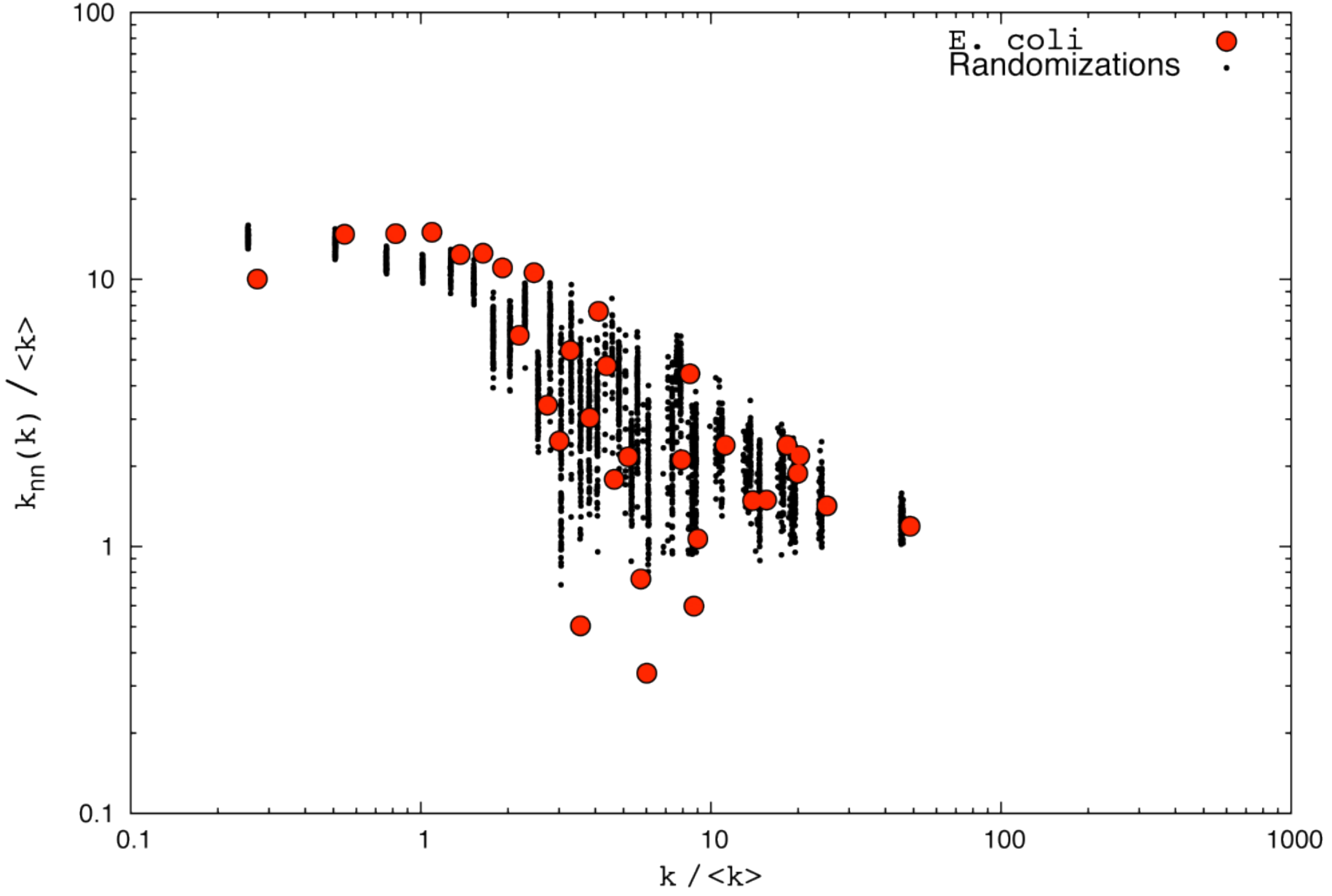}
\caption[]{{\bf Effect of randomization on the degree-degree correlation function of the {\it E. coli} 
genome.} Red discs mark the 
data for the {\it E. coli} 
GRN~\citep{RegulonDB},
superposed on the data points obtained from 
100 independent rewirings, keeping the in- and out-degree of each 
node fixed, separately.}
\label{NN-DC_random}
\end{figure}

\begin{figure}
\includegraphics[width=1.0\textwidth]{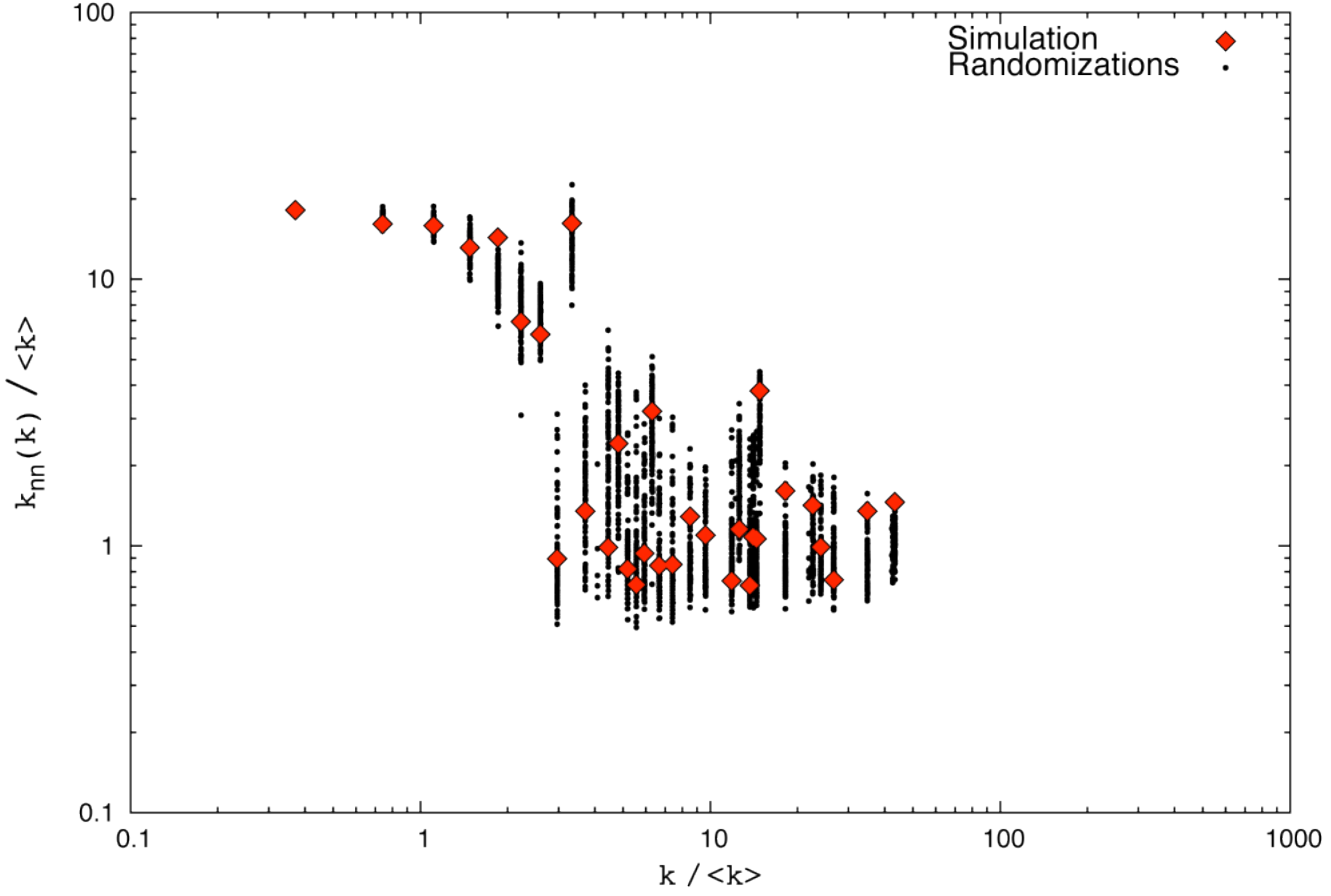}
\caption[]{{\bf Effect of randomization on the model network: degree-degree correlations.} Same as in 
Fig.~\ref{NN-DC_random}, except that instead of the 
empirical network, a randomly chosen model network (red diamonds) has been 
randomized as described there.}
\label{NN-DC_sim_random}
\end{figure}

In order to see how randomization effects the empirical and model 
gene regulatory networks (GRN) for {\it E. coli},
we display the topological 
properties of the randomized versions of the empirical (Figs. \ref{CC_random}-\ref{RCC_random}) and the 
model (Figs. \ref{RCC_sim_random}-\ref{NN-DC_sim_random}) networks. For each measured quantity, we have first 
taken the empirical network and, keeping the in- and out-degrees fixed for 
each node, randomly reconnected the edges. Next we have done the same for 
one (randomly chosen) realization of the model network. Clearly the degree 
distribution and the in- and out-degree distributions are invariant under 
this operation. The self-interactions in the randomized networks were 
removed before the calculation of the coefficients $C(k)$, $r(k)$ and 
$k_{\rm nn}(k)$.

Comparing the scatter of points in Fig. 9 with those 
for the randomized empirical network in Fig.~\ref{CC_random} shows that 
the randomized versions of the empirical network are more similar in their 
clustering coefficient to that of the set of realizations of the model 
network. Thus the rewiring has decreased the incidence of triangles in the 
empirical network towards values observed for the model network, where 
there are no correlations between the binding sequences of nodes that are 
connected to each other.  This effect is more pronounced for small $k$.
By contrast, in Fig.~\ref{CC_sim_random}, showing a set of randomized 
model networks, 
the scatter of points is evenly distributed around the original model network.

In Fig.~\ref{RCC_random}, we see that under randomization, $r(k)$ for the 
empirical network has become much more ``typical," in comparison to the ensamble of model networks in Fig. 10. However, in 
Fig.~\ref{RCC_sim_random} we observe an unexpected situation, where the 
values of $r(k)$ for the randomized model networks have systematically 
fallen below that of the original set, especially for relatively higher 
$k$ values. This indicates that there is quite a bit of variability 
between different realizations of the model network, i.e., a given 
realization can quite easily be not so ``typical."

On the other hand, the $k-k$ correlation shows a quite different behaviour 
under randomization; compare  Fig.11 with Fig.~\ref{NN-DC_random} and 
Fig.~\ref{NN-DC_sim_random}.  For both the simulations and the empirical 
network, the plot of the $k-k$ correlations of the randomized set sits more 
or less right where the original set of points are, without any marked 
shift.  The general trend shown by $k_{\rm nn}(k)$ is a direct consequence 
of the fact that the high degree nodes tend to be the TF-coding ones, 
regulating both nodes of the same type and non-TF-coding nodes, whereas 
the low degree nodes are generally those which only have in-degrees, i.e., 
are regulated by the TF coding ones. Since the in- and out-degrees of each 
node are kept fixed under the rewiring, the $k-k$ correlation function is 
essentially not affected.

\section{Statistics of the $k$-core decomposition}

\begin{figure}
\centering
\includegraphics[width=1.0\textwidth]{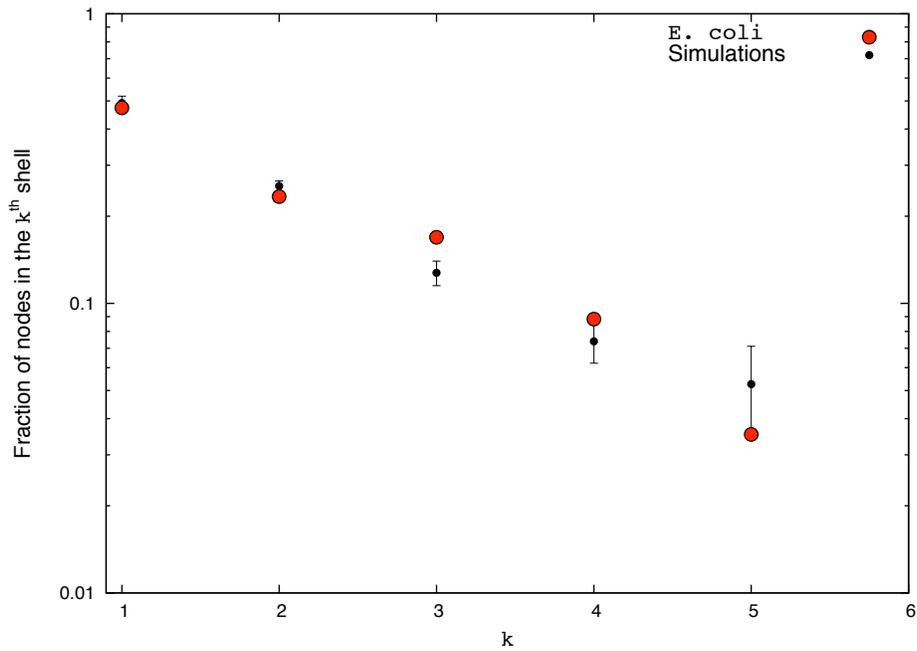}
\caption{{\bf $k$-core analysis: The coreness distribution.} We plot the relative population 
of each shell against the coreness. The red disks are for the {\it E. coli} network~\citep{RegulonDB}. For the model networks, the dependence 
on the coreness is exponential, while for the empirical network there is a small deviation especially for the shells with 
smaller coreness.}
\label{shell_size}
\end{figure}

\begin{figure}
\centering
\includegraphics[width=1.0\textwidth]{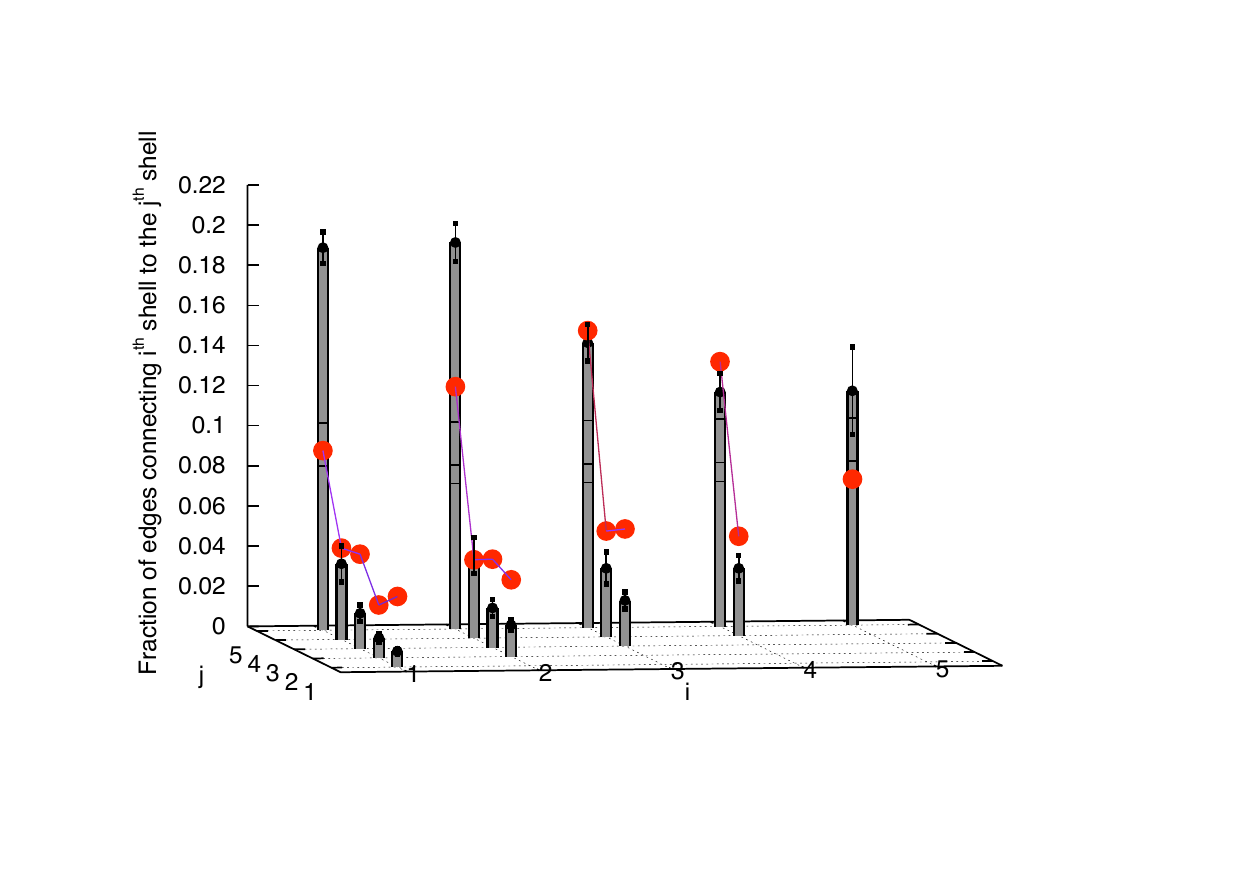}
\caption{{\bf $k$-core analysis: Edge distribution between different $k$-shells.} The fraction of edges connecting shells of coreness $i$ to shells of coreness $j$ is shown. The red dots, connected by dotted lines,  are {\it E. coli} data from the RegulonDB v6.0~\citep{RegulonDB}. Simulation results, averaged only over those  realizations with five shells, are plotted as column-graphs for better readability. The averaged values are re-plotted as black dots at the top of the columns together with error-bars corresponding to one standard deviation.   The fraction of connections to shells of higher coreness grows exponentially. The {\it E. coli} data points exhibit an excess of intra-shell connections, in comparison to the model.}
\label{edge_distribution}
\end{figure}

The $k$-core analysis of the empirical and model networks was reported in 
the main text. In Fig.~\ref{shell_size}, we plot the population of the 
shells vs the coreness, and in Fig.~\ref{edge_distribution}, the 
distribution of the number of edges 
connecting nodes within different shells. 
The distribution is very 
nearly exponential, with the empirical network deviating slighlty from the 
model one;
the shell 
population of the 
empirical network is somewhat less sensitive to the coreness.

In the statistics reported here, we have included only those realizations 
which have five shells, for ease of comparison. In both the model and the 
empirical networks, self-interactions have been removed prior to the 
analysis. 

The other network properties of the model networks with  
the number of potentially TF coding operons increased, so that those which 
actually connect match the actual number in the empirical network, 
differ very little from those reported in the main text; the 
only difference is that with higher statistics, the scatter is slightly 
reduced.  These figures are available, and will be sent electronically 
upon request from the corresponding author.

\clearpage

\begin{table}
\caption{{\bf Distribution of maximum shell number within the model ensemble.}  For 
$5.9 \%$ 
of the nodes being designated as TF coding, the average number of such nodes (in one 
set 
of 100 realizations) is 159, but the actual number of TFs establishing connections with 
binding sites turns out to be only 76 on the average. Doubling this number yields 156 
TFs which actually connect. We provide below the frequencies of model networks in either set, 
for different maximum core numbers. }
\centering
\vskip 0.5cm
\begin{tabular}{ l c  c c c c  }
\toprule
\multicolumn{5}{l}{ average number of TFs = 76.5}\\
max. core no. &2 & 3 & 4 & 5 & 6 \\
frequency& 1& 30& 52 & 14& 3\\
\multicolumn{6}{l}{$\langle k_{\rm max}\rangle =3.88$, median$= 4$}\\
\multicolumn{5}{l}{ average number of TFs = 156}\\
max. core no. &4 & 5 & 6 & 7 & 8  \\
frequency & 9 & 31& 44 & 9 &7 \\
\multicolumn{6}{l}{$\langle k_{\rm max} \rangle=5.74$, median $= 6$}\\
\bottomrule
\end{tabular}
\label{supp-comparison}
\end{table}

\clearpage

\clearpage

\end{document}